\documentclass{nature}

\newcounter{firstbib}

\usepackage{graphicx}
\usepackage{deluxetable}
\usepackage{epsfig}
\usepackage{tablefootnote}
\usepackage{url}
\usepackage{amsmath,times}
\usepackage{lineno}

\def\ra#1#2#3{#1$^{\rm h}$#2$^{\rm m}$#3$^{\rm s}$}
\def\dec#1#2#3{$#1^\circ#2'#3''$}

\newcommand{\Msun}{~M_\odot}

\newcommand{\ccm}{\rm ~cm^{-3}}
\newcommand{\kms}{\rm ~km~s^{-1}}
\newcommand{\ergs}{\rm ~erg~s^{-1}}

\newcommand{\wl}{\lambda}

\newcommand{\wll}{\lambda \lambda}

\usepackage{amssymb}
\usepackage{amsmath}
\usepackage{tablefootnote}

\newcommand{\apj}{Astrophys. J.}

\newcommand{\mnras}{Mon. Not. R. Astron. Soc.}

\title{A UV Resonance Line Echo from a Shell Around a Hydrogen-Poor Superluminous Supernova}

\author{R.~Lunnan,$^{1,2}$ 
 C.~Fransson,$^{1}$ 
 P.~M.~Vreeswijk,$^3$ 
 S.~E.~Woosley,$^4$ 
 G.~Leloudas,$^{5}$ 
 D.~A.~Perley,$^6$ 
 R.~M.~Quimby,$^{7,8}$ 
 Lin~Yan,$^9$ 
 N.~Blagorodnova,$^2$ 
 B.~D.~Bue,$^{10}$  
 S.~B.~Cenko,$^{11,12}$ 
 A.~De~Cia,$^{13}$ 
 D.~O.~Cook,$^2$
 C.~U.~Fremling,$^2$ 
 P.~Gatkine,$^{11}$ 
 A.~Gal-Yam,$^3$  
 M.~M.~Kasliwal,$^2$  
 S.~R.~Kulkarni,$^2$ 
 F.~J.~Masci,$^{14}$ 
 P.~E.~Nugent,$^{15,16}$  
 A.~Nyholm,$^1$  
 A.~Rubin,$^3$  
 N.~Suzuki,$^{8}$ and 
 P.~Wozniak$^{17}$}

\begin{document}

\maketitle

\begin{affiliations}
   \item Oskar Klein Centre, Department of Astronomy, Stockholm University, SE-106 91 Stockholm, Sweden
   \item Department of Astronomy, California Institute of Technology, 1200 East California Boulevard, Pasadena, CA 91125, USA
   \item Benoziyo Center for Astrophysics and the Helen Kimmel Center for Planetary Science, Weizmann Institute of Science, 76100 Rehovot, Israel
   \item Department of Astronomy and Astrophysics, University of California, Santa Cruz, CA 95064, USA
   \item Dark Cosmology Centre, Niels Bohr Institute, University of Copenhagen, Juliane Maries vej 30, 2100 Copenhagen, Denmark
   \item Astrophysics Research Institute, Liverpool John Moores University, IC2, Liverpool Science Park, 146 Browlow Hill, Liverpool L3 5RF, UK
   \item Department of Astronomy, San Diego State University, San Diego, CA 92182, USA
   \item Kavli IPMU (WPI), UTIAS, The University of Tokyo, Kashiwa, Chiba 277-8583, Japan
   \item Caltech Optical Observatories \& Infrared Processing and Analysis Center, California Institute of Technology, Pasadena, CA 91125, USA
   \item Jet Propulsion Laboratory, California Institute of Technology, Pasadena, CA 91109, USA
   \item Astrophysics Science Division, NASA Goddard Space Flight Center, Mail Code 661, Greenbelt, MD 20771, USA
   \item Joint Space-Science Institute, University of Maryland, College Park, MD 20742, USA
   \item European Southern Observatory,Karl-Schwarzschild Str. 2, 85748 Garching bei Munchen, Germany
   \item Infrared Processing and Analysis Center, California Institute of Technology, Pasadena, CA 91125, USA
   \item Lawrence Berkeley National Laboratory, 1 Cyclotron Road, MS 50B-4206, Berkeley, CA 94720, USA
   \item Department of Astronomy, University of California, Berkeley, CA 94720-3411
   \item Los Alamos National Laboratory, MS-D466, Los Alamos, NM 87545, USA
\end{affiliations}


\clearpage

\begin{abstract}
Hydrogen-poor superluminous supernovae (SLSN-I) are a class of rare and energetic explosions discovered in untargeted transient surveys in the past decade\cite{qkk+11,gal12}. The progenitor stars and the physical mechanism behind their large radiated energies ($\sim10^{51}$ erg) are both debated, with one class of models primarily requiring a large rotational energy\cite{kb10,woo10}, while the other requires very massive progenitors to either convert kinetic energy into radiation via interaction with circumstellar material (CSM)\cite{wbh07,ci11,mbt+13,sbn16}, or engender a pair-instability explosion\cite{gmo+09,kbl+14}. Observing the structure of the CSM around SLSN-I offers a powerful test of some scenarios, though direct observations are scarce\cite{yqo+15,ylp+17}. Here, we present a series of spectroscopic observations of the SLSN-I iPTF16eh, which reveal both absorption and time- and frequency-variable emission in the Mg~II resonance doublet. We show that these observations are naturally explained as a resonance scattering light echo from a circumstellar shell. Modeling the evolution of the emission, we find a shell radius of 0.1~pc and velocity of 3300~km~s$^{-1}$, implying the shell was ejected three decades prior to the supernova explosion. These properties match theoretical predictions of pulsational pair-instability shell ejections, and imply the progenitor had a He core mass of $\sim 50-55~\Msun$, corresponding to an initial mass of $\sim 115~\Msun$.
\end{abstract}
\begin{figure}
 \centering
\begin{tabular}{cc}
\includegraphics[width=3.5in,trim={1.0cm 1.0cm 1.0cm 1.0cm}]{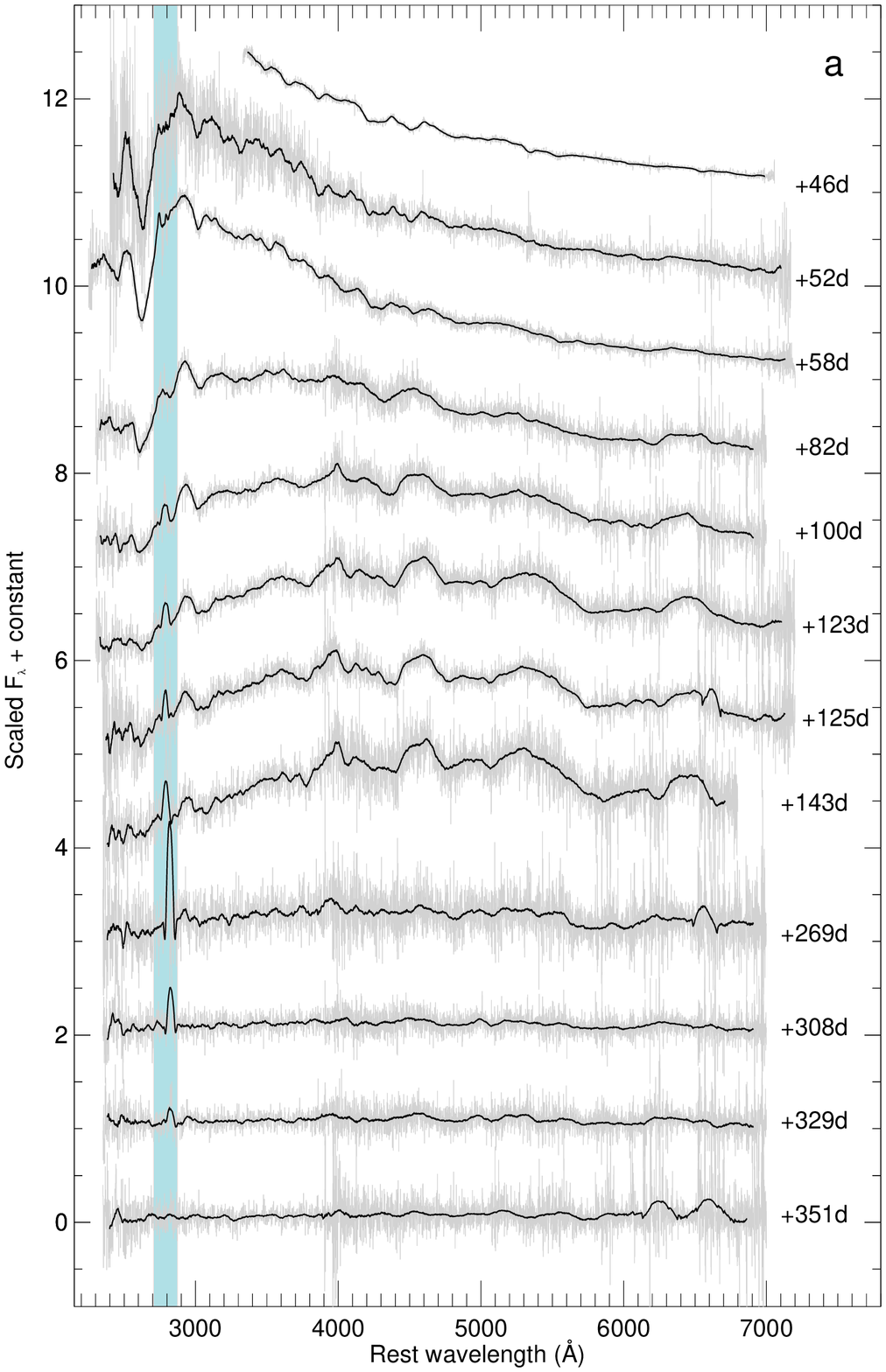} &
\includegraphics[height=6.2in,trim={0.5cm 1.0cm 1.0cm 1.0cm}]{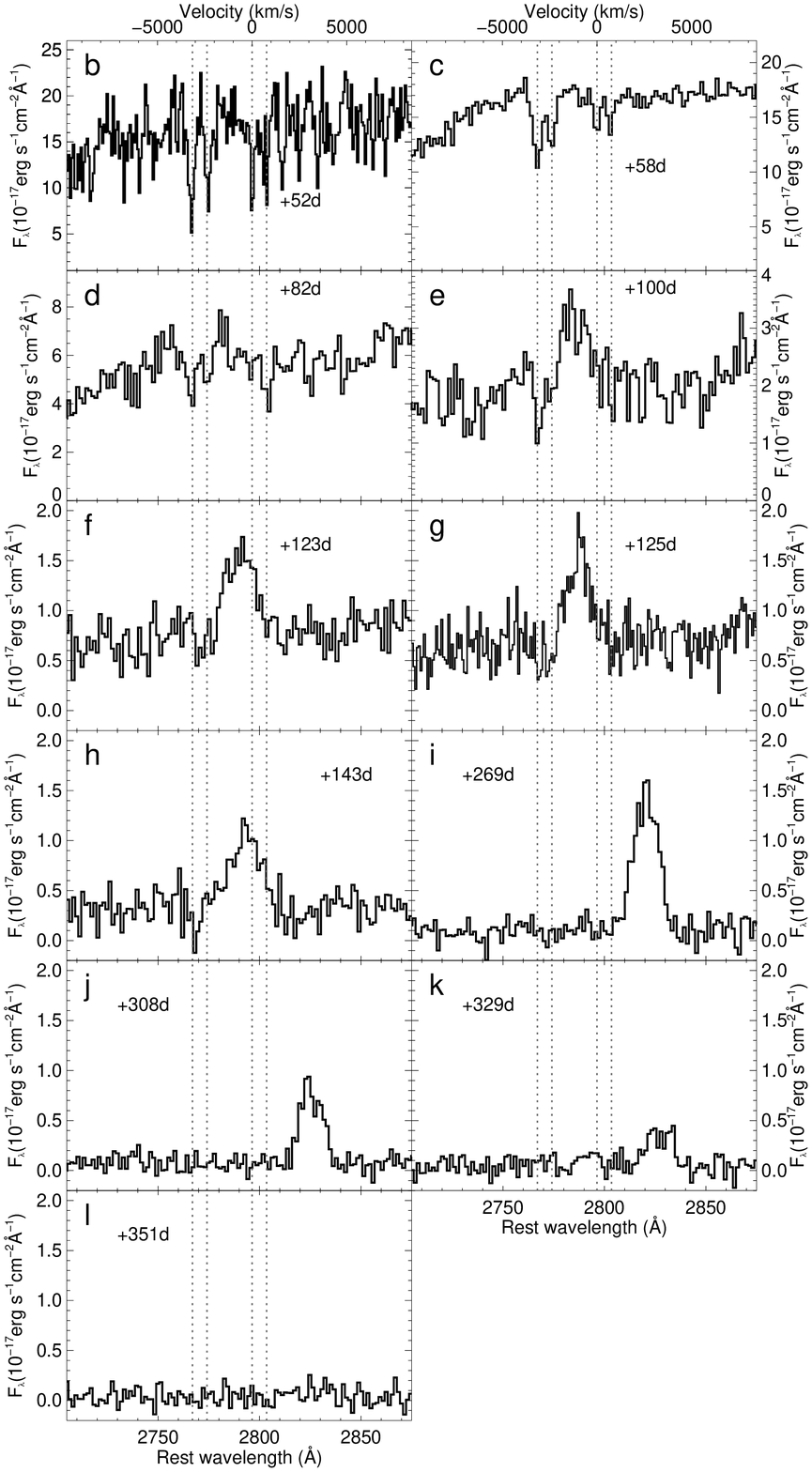}
\end{tabular}
\caption{Spectroscopic evolution of iPTF16eh. \textit{Left:} Observed spectroscopic sequence of iPTF16eh, with the phase of each spectrum in rest-frame days relative to the best-fit explosion date indicated. The gray shows the observed spectra, while the black shows the spectra smoothed by a Savitzky-Golay filter. The spectra up to day +143 have been normalized to the continuum value at 4000~\AA\,, while the spectra after 200 days have continua dominated by noise and are instead normalized by a value $5\times 10^{-18}~{\rm erg~s}^{-1}~{\rm cm}^{-2}~{\rm \AA}^{-1}$. Each spectrum is offset from each other by one scale unit. The spectroscopic evolution of iPTF16eh is similar to a typical SLSN, with the exception of a strong emission line developing around 2800~\AA. \textit{Right:} Zoom-in on the region around Mg~II (shaded blue on the left side plot), shown on an absolute flux scale and without any smoothing. The two earliest spectra clearly show the presence of two Mg~II absorption systems; the highest-redshift one is interpreted as due to absorption in the host galaxy ISM, while the second system is at slightly lower redshift, corresponding to a velocity offset of $-3200~\kms$ with respect to the host galaxy. Note the development of an emission line around 100~days past explosion, the redward shift of the emission line with time, and its fading away at 330-350~days after explosion.
  \label{fig:spec}}
 \end{figure}



iPTF16eh was first detected by the intermediate Palomar Transient Factory\cite{lkd+09} on 2015 December 17.5 (UT). An initial spectrum, taken on 2016 February 18 showed a blue continuum with shallow O~II absorptions characteristic of SLSN-I (Figure~\ref{fig:spec}); subsequent spectra with bluer wavelength coverage revealed two sets of narrow Mg~II absorption lines. Assuming the higher-redshift absorption lines originate in the interstellar medium of the host galaxy, the redshift of iPTF16eh is $z=0.427$. We continued to follow the evolution of iPTF16eh photometrically and spectroscopically, and find that it is among the most luminous SLSNe discovered to date with a peak absolute magnitude of $M_u = -22.55$~mag (AB)\cite{lcb+17,dgr+17}, but otherwise evolves similarly to other SLSN-I with slow timescales. However, it displays one unusual feature never seen before in SLSN spectra: an intermediate-width emission feature around 2800~\AA\, in the rest frame, also likely associated with Mg~II, emerging around 100~days after explosion (taking the explosion date to be 2015 December 14.5; see Methods). 

\begin{figure}
\centering
\includegraphics[width=3.5in]{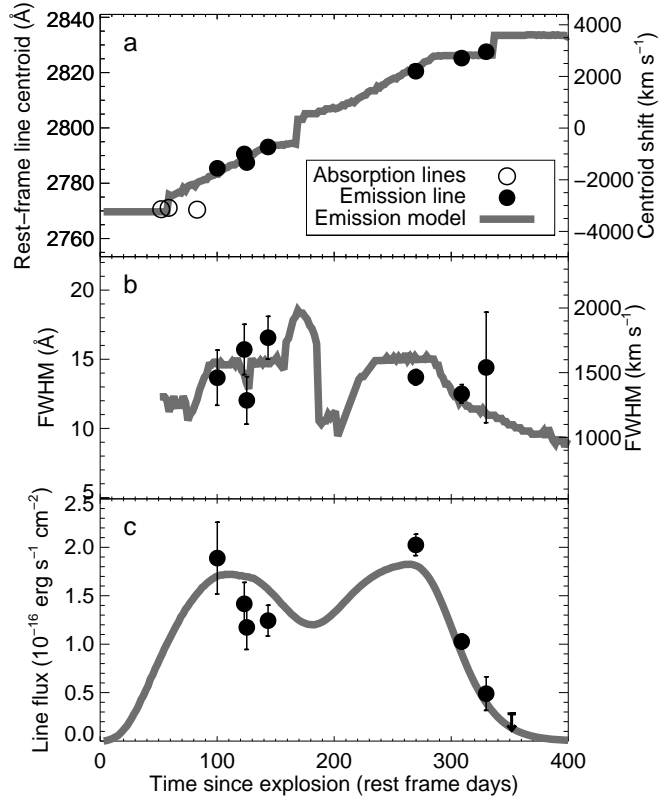}
\caption{Observed and modeled evolution of the Mg~II emission line. The black filled circles show measured line properties as measured by fitting a Gaussian profile, using the {\tt lmfit} routine in IDL; the plotted data is also tabulated in Supplemental~Table~3. The error bars plotted are the 1$\sigma$ estimated uncertainties on each parameter, given the measurement errors of the input spectra. The gray line shows the corresponding line evolution in the resonance scattering model. 
\textit{(a)} Line centroid versus time. In addition to the filled circles showing the evolution of the emission line, the open circles show the average of the absorption line doublet, where measurable. Typical error bars are $\lesssim 1$~\AA\, and are too small to show on the plot. \textit{(b)} Line FWHM versus time. The jumps in the model at $\sim$ 200 days are caused by the Mg II absorption lines from the host galaxy. \textit{(c)} Line flux versus time. 
\label{fig:mgii_prop}}
\end{figure}

The right panel of Figure~\ref{fig:spec} shows the development of the Mg~II emission line in relation to the two Mg~II absorption systems. The line is first clearly visible in the spectrum taken 100~days past explosion, and persists for more than 200~days, finally fading below detectability 350~days after explosion. To quantify the properties of the line, we fit a Gaussian profile, and measure the line centroid, flux, and full-width half maximum (FWHM) as a function of time (Fig.~\ref{fig:mgii_prop}). The centroid of the line clearly shifts redwards with time; initially the emission line is blueshifted with respect to the supernova redshift by $\sim 1600~{\rm km~s}^{-1}$, but reaches a redshift of $\sim 2900~{\rm km~s}^{-1}$ before fading away. The line flux varies at most by a factor of two until the line starts fading, and the FWHM of the line similarly remains approximately constant around $\sim 1500~{\rm km~s}^{-1}$.

Both the blueshifted absorption system and the existence and properties of the emission line can be naturally explained by resonance line scattering of the SLSN continuum by a rapidly expanding, roughly spherical CSM shell. Mg~II ions in the shell absorb continuum photons around 2800~\AA\, from the SLSN, which we see along our line of sight as a blueshifted absorption feature in the SLSN spectrum. These excited Mg~II ions then almost instantaneously (within $10^{-8}$~s) decay back down to the ground level, and emit a line emission photon in a random direction, which we observe as an emission line. Because the light travel time across the shell ($R_{\rm out} / c$) is longer than the duration of the SLSN light causing the excitation, we see different parts of the shell light up at different times (Supplementary Fig.~5), explaining the drift of the emission line from blueshifted to redshifted as we go from seeing the front to the back of the shell.

To quantify the shell properties, we have done Monte-Carlo calculations of the scattering process (Methods), assuming a spherical shell with inner radius $R_{\rm in}$ and outer radius $R_{\rm out}$. The size of the shell is constrained by the duration of the emission, while the thickness is determined by the relative intensity of the emission lines and the scattering continuum, $F^{\rm line}_{\nu}/F^{\rm cont}_{\nu}(t_{\rm peak}) \approx \Delta R/R$. We assume homologous expansion with $V = V_{\rm max} (r/R_{\rm out}) \kms$ for the shell, which is likely for a time-limited eruption, like that resulting from the 1843 eruption in Eta Carinae\cite{mkb+01}, or for a pulsational pair-instability ejection\cite{wbh07}; however, the results are not sensitive to this assumption.
In addition to the parameters describing the shell, the resulting emission depends on the light curve of the scattered radiation, i.e. the supernova continuum around 2800~\AA. As our observed photometry does not extend sufficiently to the blue, we measure the continuum in this wavelength region from the observed spectra (Supplementary~Fig.~1). Unfortunately, our earliest spectrum covering this wavelength region is 87~days after explosion, so we need to make assumptions for the rising light curve at this wavelength; prior to the first spectrum, we therefore use a stretched form of the modelled $u$-band light curve of PTF12dam\cite{tnb+17,nsj+13}, which fits the observed 2800 \AA \ light curve after the peak (Methods).

\begin{figure}
\centering
 \includegraphics[width=5in]{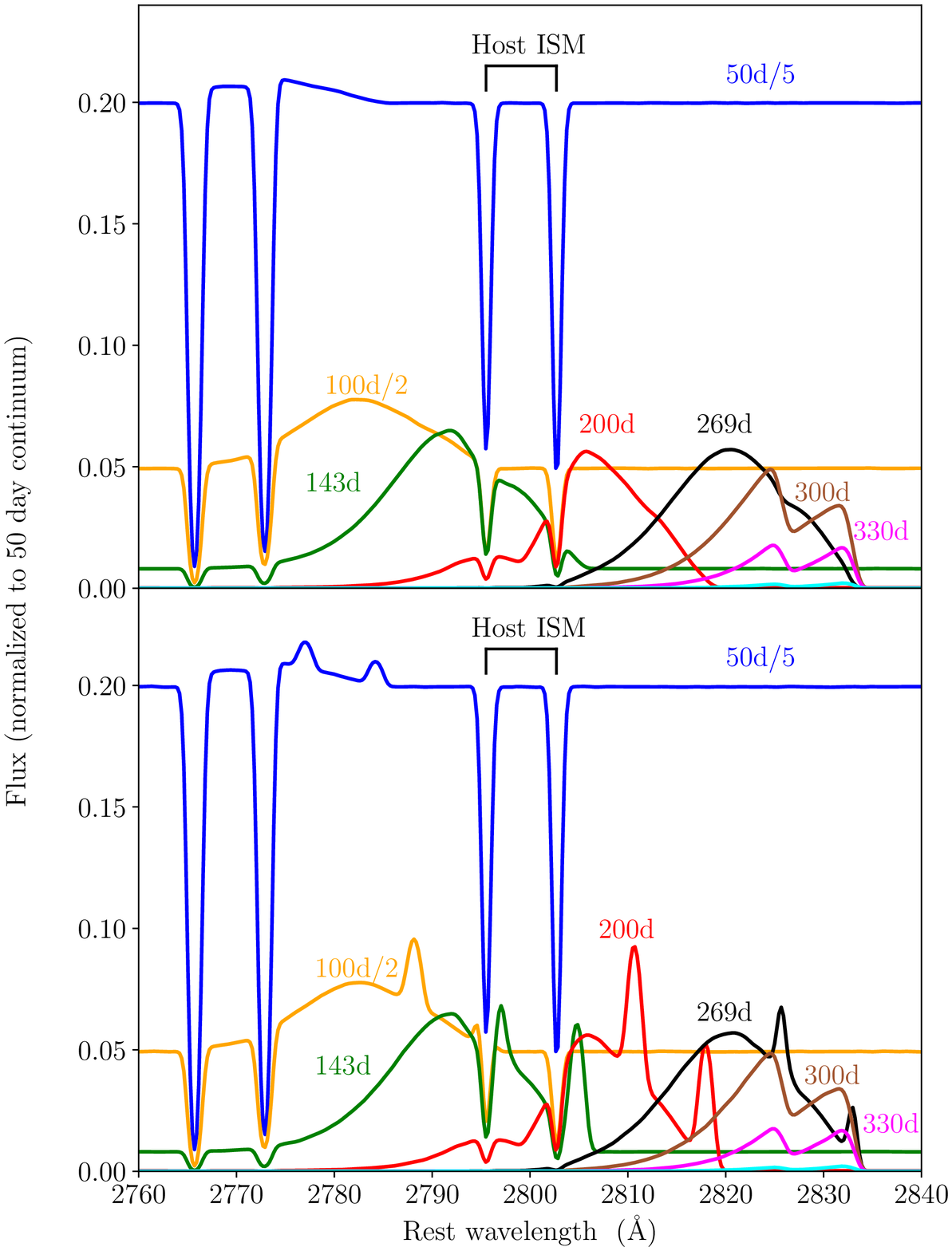}
\caption{Time sequence of simulated spectra in the Mg~II wavelength region for a shell with inner radius 130 light days and outer radius 137 light days for two different light curves for the 2800 \ \AA \ flux. \textit{Upper panel:} Our standard model. The flux of each spectrum has been scaled by a common factor and normalized to the continuum level at 2800 \ \AA at 50 days; the two first spectra have then been scaled down by a further factor of 5 and 2, respectively (as indicated) to facilitate comparison.
\textit{Bottom panel:} Same as the upper panel, but with a luminous burst added at 1 day. We have added absorption lines from the host galaxy and convolved the spectrum with the instrumental resolution of the Keck/LRIS spectrograph with the grating used (FWHM $ \approx 6.5$ \ \AA). 
\label{fig:model1}}
\end{figure}

Fig.~\ref{fig:model1} shows the line profiles at the observed epochs for $R_{\rm in}=130$ light days and  $R_{\rm out}=137$ light days and $V_{\rm max}=3300 \kms$, and in Fig.~\ref{fig:mgii_prop} the centroid and flux from the model. As is seen, there is a good qualitative agreement with the observed evolution in Figs.~\ref{fig:spec} and \ref{fig:mgii_prop}. In particular, the fading of the line flux gives a strong constraint on the outer radius of the shell, and we estimate the error to be $\pm 7$ light days (Methods).
In the first two epochs, the SN continuum dominates the light output, and we see strong absorption together with a weak, but growing, emission component on top of the SN spectrum. The spectrum at 143 days marks the transition from an absorption dominated to a pure emission spectrum, as the direct continuum from the SN becomes too faint to show any absorption component. The flux of the simulated emission lines is nearly constant up to $\sim 270$ days, when it drops sharply. This is in good agreement with the observations (Fig.~\ref{fig:mgii_prop}), and corresponds to the constant time parabola of the echo exiting from the outer radius at the far side of the shell (Supplementary~Fig.~5). At 330 days only weak emission remains as the light from the supernova tail is scattered off the back of the shell.

The line shapes are sensitive to the input light curve, and can therefore serve as a probe of the (unobserved) early supernova light curve around 2800~\AA. To explore this, we also run a model including a luminous shock breakout, parametrized as a burst of radiation at 1 day and a Gaussian shape with $\sigma = 1$ day and a peak luminosity a factor of 5 brighter than that of the main peak at the relevant wavelengths around 2800~\AA. While we do not explicitly assume a temperature for this shock breakout contribution, it is implied in the model that a sufficiently small amount of the Mg~II is ionized to leave the line optically thick.
The bottom panel in Fig.~\ref{fig:model1} shows the result of this calculation, which leads to two distinct peaks at the long wavelength edge of the line. The width of the line peaks are mainly determined by the width of the shell, as long as the burst is short compared to the width of the shell. These examples show that the late line shape serves as a powerful diagnostic of the early light curve, and in particular that in the case of iPTF16eh the echo is consistent with being dominated by the main SN light curve, as opposed to a strong, brief shock breakout pulse on day one.

We can also place constraints on the geometry of the shell. The smooth velocity evolution, as well as the very similar velocity of absorption ($-3,200 \kms$) and maximum emission ($\sim 3,400 \kms$) argue for a symmetric eruption. The fact that both absorption and emission lines are seen also argues against a thin ring (similar to what was seen in SN\,1987A), since this would require the plane of the ring to be very close to the line of sight. Moreover, we find that the total scattered flux in the absorption components is consistent within the errors of the flux in the integrated emission (Methods), suggesting that the covering factor of the shell is close to $4\pi$ steradians.  As discussed in the Methods, the linear  evolution of the line velocity indicates that the geometry is also close to spherical. This therefore argues against e.g., a bipolar structure such as that of Eta Car, and may also indicate that rotation of the progenitor was not important for the ejection.

As Mg~II is the only line we detect from the shell, we can only place weak constraints on the composition, and therefore the total mass of the shell. The nondetection of other lines does not imply an unusual composition, as Mg~II is the strongest resonance line of abundant elements in the observed part of the spectrum, and other possible species like Na~I, K~I and Ca~II are likely ionized by the hot supernova continuum. The Mg~II doublet is at least partially saturated, so the equivalent width of the Mg~II$\lambda 2803$ line only allows us to set a lower limit on the column density of $N_{\rm Mg~II} \gtrsim 10^{14}$~atoms~cm$^{-2}$. The lack of detected H$\alpha$ emission from the shell does not necessarily imply that the shell is H-poor, but places an upper limit on the H mass of $M \lesssim 27 f \Msun$, where $f < 1$ is the filling factor of the shell (Methods).

Given the derived size of $\sim 3.5 \times 10^{17}$~cm and observed velocity of $\sim 3300~\kms$, the shell was ejected $\sim 32$~years prior to the supernova explosion, assuming a constant velocity. We consider several different mechanisms for the origin of the shell.
In the Methods we rule out the wind of a Wolf-Rayet progenitor.
A more likely possibility is that the shell could be the result of a previous LBV-like massive ejection, similar to that seen in Eta Carinae.
In particular, high velocity material moving up to $\sim 6000~\kms$  was seen in the 1840 eruption of Eta Carina\cite{smi08}.
In addition, the 2012A outburst of SN\,2009ip shows that ejections with velocities of $\gtrsim 6000 \kms$ can occur without disrupting the star\cite{pci+13,msf+13}. A problem for the LBV eruption scenario is that the structure of the CSM may be highly anisotropic with material moving at a range of velocities, as is seen for Eta Car\cite{smi08}, unlike the detached shell we see in iPTF16eh.

Our preferred alternative is that the shell is the result of a pulsational pair-instability (PPI) ejection, which has been discussed in connection to both H-rich and H-poor SLSNe\cite{qkk+11,wbh07,sbn16,tnb+17,woo17,yqo+15,ylp+17}. The time scale between the first major ejection and the final collapse is a strong function of the He core mass; 32 years would correspond to a He core mass of $\sim 51-53 \Msun$  according to Ref.~22\nocite{woo17}. Depending on the amount of rotation this implies a zero-age main sequence mass of $\sim 90 - 120 \Msun$. Ejection of $\lesssim 10 \Msun$ of material in the initial pulse results in velocities $\sim 2700-2900 \kms$ (Ref.~22), close to our observed velocities, and thus shows that one can get a consistent picture in this scenario. Subsequent pulses happen closer to the final collapse for cores in this mass range ($\sim 30$~days before explosion in the $53~\Msun$ model), and would be already swept up by the supernova ejecta by the time of our first spectrum. Better sampled light curve data as well as spectroscopy during the rising part of the light curve could constrain the presence of such later-ejected shells and offer a test of the PPI scenario.

Although our observations are consistent with a shell resulting from a PPI ejection, this does not imply that the supernova light curve is also explained by interaction between later shell ejections. In particular, the total energy and luminosity expected from such shell ejections are lower than what is observed for iPTF16eh by about an order of magnitude\cite{woo17}.
Possibilities, discussed in Methods, include magnetars in both lower mass stars and high mass stars, ending their lives as SNe. A further possibility is energy injection from a  disk around a black hole, resulting from a rapidly rotating massive progenitor. 
Whatever mechanism may be driving the main supernova light curve, our observations put strong constraints on any progenitor model by requiring a significant mass ejection during the final burning stages before explosion.

Ultimately the supernova ejecta will collide with the shell, which may provide a chance to estimate the composition from the emission. This will occur at a time $R_{\rm in}/V_{\rm ejecta}$.
From the spectra, we measure a maximum ejecta velocity of $\sim 15,000~\kms$ (Methods), and thus predict that the collision will take place $\sim 7$~years after the supernova explosion (in the rest frame), or 10~years as observed on Earth. This will result in optical/UV, radio and X-ray radiation, and may in principle be observable depending on the mass of the shell. Especially monitoring the optical flux may be promising for information about the chemical composition of the shell: if there is hydrogen present, we may expect H$\alpha$ emission, which has indeed been seen in other SLSN-I with late-time CSM interaction\cite{yqo+15,ylp+17}.

Finally, we consider how unique iPTF16eh is. At least two favorable conditions aligned to make the resonance echo observable: first, the supernova redshift is sufficiently high ($z \gtrsim 0.25$) for Mg~II to be easily observable by ground-based optical spectrographs, which typically have limited sensitivity below 3500-4000~\AA\,. In addition, the supernova itself is among the most luminous ever discovered, making it possible to obtain high-quality spectroscopic observations over a long time baseline. Higher-redshift SLSNe that cover the rest-frame UV rarely have late-time spectra available, and thus would not be able to detect a Mg~II emission line like the one seen in iPTF16eh. However, the absorption lines from the shell would be readily visible in spectra taken at peak brightness. We have searched through all the spectra of SLSNe from the Pan-STARRS Medium Deep Survey\cite{lcb+17} as well as PTF/iPTF\cite{dgr+17,qdg+18}, and we do not find evidence of a double Mg~II absorption system in any of the spectra with sufficient quality blue spectra to cover these wavelengths (27 total). Given that in the PPI model the time between the first shell ejection and the final core-collapse is a strong function of the He core mass, it is possible that even if such ejections are common, a shell at the distance and velocity separation of iPTF16eh could be rare. The recent detections of late-time H$\alpha$ emission in three SLSN-I\cite{yqo+15,ylp+17}, interpreted as collision with circumstellar shells located at distances $\sim 10^{16}$~cm, suggest that iPTF16eh is not alone in having a complex circumstellar environment, although the detection of the echo gives unique information about the location, geometry, velocity and mass loss time scale.

\clearpage



 \clearpage
 
 \begin{addendum}

\item[Correspondence] Correspondence and requests for materials
should be addressed to Ragnhild Lunnan~(email: ragnhild.lunnan@astro.su.se).

\item 
We are grateful for discussions with Claes-Ingvar Bj\"ornsson, Sergei Blinnikov, Elena Sorokina, Enrico Ramirez-Ruiz and Jim Fuller. We are also grateful to the referees for several clarifying comments.  The Intermediate Palomar Transient Factory project is a scientific collaboration among the California Institute of Technology, Los Alamos National Laboratory, the University of Wisconsin, Milwaukee, the Oskar Klein Centre, the Weizmann Institute of Science, the TANGO Program of the University System of Taiwan, and the Kavli Institute for the Physics and Mathematics of the Universe.
This work was supported by the GROWTH project funded by the National Science Foundation under Grant No 1545949.
This research was supported by the Swedish Research Council, the Swedish National Space Board, and the Knut and Alice Wallenberg Foundation.
Part of this research was carried out at the Jet Propulsion Laboratory, California Institute of Technology, under a contract with the National Aeronautics and Space Administration. 
A.G.-Y. is supported by the EU via ERC grant No. 725161, the Quantum Universe I-Core program, the ISF, the BSF Transformative program and by a Kimmel award.
PEN acknowledges support from the DOE through DE-FOA-0001088, Analytical Modeling for Extreme-Scale Computing Environments.
This research used resources of the National Energy Research Scientific Computing Center, a DOE Office of Science User Facility supported
by the Office of Science of the U.S. Department of Energy under Contract No. DE-AC02-05CH11231.
IRAF is distributed by the National Optical Astronomy Observatory, which is operated by the Association of Universities for Research in Astronomy (AURA) under cooperative agreement with the National Science Foundation.
Some of the data presented herein were obtained at the W.M. Keck Observatory, which is operated as a scientific partnership among the California Institute of Technology, the University of California and the National Aeronautics and Space Administration. The Observatory was made possible by the generous financial support of the W.M. Keck Foundation. The authors wish to recognize and acknowledge the very significant cultural role and reverence that the summit of Mauna Kea has always had within the indigenous Hawaiian community.  We are most fortunate to have the opportunity to conduct observations from this mountain.

\item[Author Contributions] 
RL coordinated the observational campaign, was PI of the Keck program under which the late-time spectra were obtained, analyzed the data, and wrote the manuscript.
CF wrote the resonance scattering code, ran the simulations, performed model comparisons, and contributed to manuscript writing.
PMV contributed to the interpretation and resonance line calculations and manuscript preparation.
SEW contributed to the comparison with PPI models and manuscript preparation.
GL, DAP, RMQ, LY, ADC, AG-Y contributed to the discovery, analysis, interpretation and manuscript preparation.
MMK contributed to manuscript preparation.
SRK is the PI of iPTF and of the P200/Keck programs under which the early spectra were taken, and contributed to manuscript preparation.
AN contributed to finding the supernova and to manuscript preparation.
NB, DOC and AR reduced spectra.
SBC and PG obtained and reduced DCT photometry.
CUF reduced the P60 photometry. 
NS obtained and reduced the Subaru spectrum.
FJM and PEN contributed to the photometric pipelines applied by iPTF.
BDB and PW contributed to the iPTF machine learning codes for transient search.


 \item[Competing Interests] The authors declare that they have no
competing financial interests.

\end{addendum}


 \clearpage

 \clearpage
 
\clearpage

\begin{methods}

\subsection{iPTF discovery and classification}
iPTF16eh was first detected by the intermediate Palomar Transient Factory at coordinates RA=\ra{12}{41}{06.21}, Dec=\dec{+32}{48}{30.9} (J2000) on 2015 December 17.5 (UT dates are used throughout this paper), at a magnitude $g = 21.43 \pm 0.23~{\rm mag}$. The photometric coverage on the rise is sparse, as iPTF was conducting a ``slow-and-wide'' experiment at the time; iPTF16eh was eventually saved and flagged for follow-up on 2016 February 8. A spectrum taken with the Faint Object Camera and Spectrograph (FOCAS\cite{kaa+02}) on the 8.2~m Subaru Telescope on 2016 February 18 revealed a blue continuum with O~II absorptions, classifying iPTF16eh as a SLSN-I at an approximate redshift of $z \simeq 0.42$. A subsequent spectrum taken with the Double Beam Spectrograph (DBSP\cite{og82}) on the 200-in Hale telescope at Palomar Observatory on 2016 February 27 shows narrow Mg~II~$\lambda\lambda$2796,2803 absorption at $z=0.413$, though a better S/N spectrum taken with Low Resolution Imaging Spectrometer (LRIS\cite{occ+95}) on the 10-m Keck I telescope on 2016 March 6 also reveals the presence of a second, weaker Mg~II$\lambda\lambda$2796,2803 absorption system at $z=0.427$. We take this higher-redshift system to be the redshift of the supernova host galaxy.

\subsection{Photometry}
The $g$-band photometry obtained with the P48 CFH12K camera was processed with the Palomar Transient Factory Image Differencing and Extraction (PTFIDE) pipeline\cite{mlr+17} to obtain point-spread function (PSF) photometry. In addition, we obtained $Bgri$ photometry with the automated 60-inch telescope at Palomar(P60;\cite{cfm+06}), including $gri$ data taken with the Spectral Energy Distribution Machine (SEDM\cite{bnw+17}), and PSF photometry was performed using {\tt FPipe}\cite{fst+16}. Additional $gri$ imaging was obtained with the Large Monolithic Imager (LMI) mounted on the 4.3\,m Discovery Channel Telescope (DCT) in Happy Jack, AZ. Standard CCD reduction techniques (e.g., bias subtraction, flat fielding) were applied using a custom IRAF pipeline.  Individual exposures were astrometrically aligned with respect to reference stars from the Sloan Digital Sky Survey\cite{sdssdr10} (SDSS) using SCAMP\cite{ber06}.  We calculated aperture photometry magnitudes for the transient using an inclusion radius matched to the FWHM of the PSF, and calibrated the images with respect to point sources from SDSS. Additional $gr$ images were also obtained with LRIS on the 10-m Keck~I telescope, and were reduced using {\tt LPipe} (\url{http://www.astro.caltech.edu/~dperley/programs/lpipe.html}). Again, aperture photometry was performed and calibrated against point sources in SDSS. 
iPTF16eh was also observed for one epoch with the Ultra-Violet/Optical Telescope (UVOT) aboard the Neil Gehrels \textit{Swift} Observatory; photometry was performed using the {\tt HEAsoft} package (\url{https://heasarc.nasa.gov/lheasoft/}).
Supplementary~Fig.~1 shows the observed $Bgri$ light curves, and all photometry is listed in Supplementary~Table~1. All photometry has been corrected for Milky Way foreground extinction according to $E(B-V) = 0.015$\cite{sf11}. We assume a standard $\Lambda$CDM cosmology with $\Omega_{\rm M} = 0.27$, $\Omega_{\Lambda} = 0.73$ and $H_0 = 70$~km~s$^{-1}$\cite{ksd+11}. 

In addition to the limits from the supernova discovery observing season listed in Supplementary~Table~1, the field of iPTF16eh was observed by PTF and iPTF a number of times prior to the discovery of the supernova. Pre-explosion limits exist for the following date ranges: May 13-19 and July 5, 2009; March 18 - June 13, 2010; March 1-2, 2011; February 1 - June 20, 2013; March 19 - May 28 and December 20, 2014; and June 5-26, 2015. All upper limits are in $r$-band except for March 1-2, 2011 and June 5-26, 2015 which are $g$-band. Typical nightly upper limits are between 20.5 - 21.5~mag.

\subsection{Spectroscopy}
We obtained a total of 12 spectra of iPTF16eh, taken with the FOCAS on the 8.2~m Subaru Telescope, DBSP on the 200-in Hale Telescope at Palomar Observatory, and LRIS on Keck I. LRIS spectra were reduced using LPipe, and FOCAS and DBSP spectra were reduced using standard pipelines. Supplementary~Table~2 summarizes the spectroscopic observations, and the sequence of spectra is shown in Fig.~\ref{fig:spec}. All spectra will be made publicly available through the Weizmann Interactive Supernova Data Repository (WISeREP\cite{yg12}).

\subsection{Supernova properties}
We constrain the explosion date from the $g$-band light curve, which is the only filter in which we sample the rise. Fitting a second-order polynomial (in flux space) to the rising and peak portion of the observed $g$-band light curve, we find a best-fit explosion date of 2015 December 14.5 (MJD 57370.5) $\pm 0.7~{\rm day}$, suggesting that the supernova was discovered just a few days after explosion. Similarly, we find the peak in $g$-band to be on 2016 February 6.3 (MJD 57424.3) $\pm 0.8~{\rm day}$. This gives a rest-frame rise time of $\sim 38$~days. The best-fit peak $g$-band magnitude is $19.01$~mag; given the redshift, observed $g$-band is close to rest-frame $u$-band, and we calculate a peak absolute magnitude of $M_u = -22.55$~mag (including a cross-filter K-correction, calculated using SNAKE\cite{isg+16}). This makes iPTF16eh one of the most luminous SLSNe observed to date, comparable to e.g. iPTF13ajg\cite{vsg+14}, Gaia16apd\cite{kbm+17} and PS1-13or\cite{lcb+17}.

From peak light onwards we have multi-filter data, which we can use to measure other supernova properties such as the temperature evolution and pseudo-bolometric flux. Fitting a blackbody, we find that iPTF16eh had a color temperature of $\sim 16,000~{\rm K}$ near peak, and declined at a rate of approximately $110~{\rm K~day}^{-1}$ post-peak. We follow the procedure outlined in\cite{lcb+17} to construct a pseudo-bolometric light curve, and find a lower limit on the peak bolometric luminosity of $2.7 \times 10^{44}~{\rm erg~s}^{-1}$, and a minimum radiated energy of $1.3 \times 10^{51}$~erg. These are, again, at the high end of what has been observed for SLSNe, though within the observed distribution\cite{dgr+17,lcb+17}. The light curve timescales are similar to other ``slowly-evolving'' SLSNe such as PTF12dam and PS1-11ap\cite{nsj+15}. The blackbody temperatures and radii, as well as our pseudo-bolometric light curve (assuming a constant bolometric correction on the rise) are shown in Supplementary~Fig.~2, and listed in Supplementary~Table~4.

Given that iPTF16eh is the only SLSN-I with a detected light echo, it is interesting to consider how its overall properties compare to other SLSN-I. One statistical way to describe SLSN-I is by using the peak luminosity, decline rate over 30 days, and colors at peak and after 30 days\cite{ipg+18} in two standard filters centered at 400~nm and 520~nm, respectively. We calculate $K$-corrections from our spectra to these two filters using SNAKE\cite{isg+16}. We use our earliest spectrum (taken 8.2~days past the $g$-band peak) for $K$-corrections at peak, and interpolate between the corrections calculated from the spectrum at 20.2~days and 44.8~days to get approximate $K$-corrections 30~days past peak. We find values of $M(400)_0=-22.3 \pm 0.1~{\rm mag}$, $\Delta M(400)_{30} = 0.2 \pm 0.1~{\rm mag}$, $M(400)_0-M(520)_0 = -0.26 \pm 0.1~{\rm mag}$, and $M(400)_{30}-M(520)_{30} = -0.02 \pm 0.05~{\rm mag}$. These place iPTF16eh well within the main population of SLSN-I as defined in \cite{ipg+18}, with particularly similar parameters to iPTF13ajg\cite{vsg+14}.

We note that over the time period where we can calculate $K$-corrected photometry and colors (out to $\sim 140$~days past explosion) the rest-frame colors of iPTF16eh are similar to other slowly-evolving SLSN-I with data at such epochs, such as SN\,2015bn\cite{nbs+16}; this should not be surprising given the similarity of the spectra at these epochs (Supplementary Fig.~3). In particular, we see no evidence of an excess blue continuum such as one might expect from a traditional scattered light echo by dust in the shell (if dust is present and of sufficient density). At later epochs, there is not sufficient signal in the continuum to constrain or compare the rest-frame colors. 

The overall spectroscopic evolution of iPTF16eh is also very similar to other well-studied H-poor SLSNe -- Supplementary~Fig.~3 shows comparisons to iPTF13ajg\cite{vsg+14}, PTF12dam\cite{nsj+13} and SN\,2015bn\cite{nbs+16} at similar epochs. Near peak, the spectrum shows a blue continuum with weak O~II features in the optical, with a broad, stronger UV feature bluewards of 2800~\AA\, that has been interpreted as a blend of Mg~II and C~II\cite{hkl+13}. As the ejecta cool, the spectrum evolves to show features typical of stripped-envelope SNe, mainly from Fe~II, Ca~II, Si~II and Mg~II. 
It has been shown that SLSN-I can be defined based on their spectroscopic properties alone\cite{qdg+18}, and indeed we find that the spectra of iPTF16eh fall in the SLSN-I group when applying the same template-matching framework. Additionally, it falls into the 'PTF12dam-like' subgroup rather than the 'SN\,2011ke-like' one, as may also be expected by the slow light curve evolution and the spectroscopic similarity to other slow-evolving SLSN-I. As can be seen from the comparisons with the other SLSNe, apart from the intermediate-width Mg~II emission line at late epochs discussed in the main text, iPTF16eh does not display any unusual spectroscopic features. 

Determining the ejecta velocity from SLSN spectra can be challenging, as the lines available at peak light are different from those most often used and studied in other stripped-envelope SNe (e.g., Fe~II$\lambda$5169 or Si~II$\lambda$6355). From the ``W''-shaped O~II feature around 4300~\AA\, in our earliest spectrum (Fig.~\ref{fig:spec}), we derive an ejecta velocity of $8,500~\kms$. On the other hand, if we assume that the broad UV feature at $\sim 2620$~\AA\, is dominated by Mg~II, we would derive a much higher velocity of $\sim 20,000~\kms$. This discrepancy could be due to the Mg~II line forming further out in the ejecta than the O~II line, or it could indicate that other ions than Mg~II (e.g. C~II) contribute significantly to the line shape. The evolution of the blackbody radius (Supplementary~Fig.~2) suggests an expansion velocity of $\sim 8,000~\kms$, consistent with the measurement from the O~II lines. The blue edge of the same absorption line suggests there is material moving at speeds up to at least $15,000~\kms$.

\subsection{Host Galaxy}
No host galaxy is detected at the location of iPTF16eh in either SDSS or Pan-STARRS1 3$\pi$ pre-explosion images, and our latest photometry measure a combined supernova and host brightness $m_r \simeq 25.2 \pm 0.2~{\rm mag}$. This places an upper limit on the host absolute magnitude of $M \gtrsim -16.2~{\rm mag}$ at an effective wavelength of $\sim 4300$~\AA\,, suggesting that the host of iPTF16eh is a faint dwarf galaxy. Such an environment is not unusual for SLSN-I, which show a strong preference for low-mass, low-metallicity host galaxies\cite{lcb+14,pqy+16,csy+17}. 

We measure the equivalent width of the Mg~II lines from the host galaxy ISM to be about $1.9 \pm 0.4$~\AA\,, calculated as the average from the spectra at +52 and +58 days. The main source of uncertainty comes from the uncertainty in the continuum level (from both measurement uncertainty, and the difficulty of placing the continuum given the adjacent features in the supernova spectrum). This value is consistent with the distribution for SLSN-I host galaxies compiled in\cite{vsg+14}, and again suggests that the host galaxy of iPTF16eh is not unusual in the context of SLSN-I host galaxies.

\subsection{Modeling the Scattered Shell Emission}

We calculate synthetic spectra using a Monte Carlo code, which calculates the scattering by the shell, assuming optically thick lines and isotropic scattering in the rest frame of the shell. The basic geometry is shown in Supplementary~Fig.~5, along with light echo parabolas showing which part of the shell is contributing to the observed emission at three different epochs. The outer parabola corresponds to the time of supernova shock breakout, while the inner parabola corresponds to the fading away of the supernova light. As illustrated, the observed emission line profile from the shell relates to both the input light curve and the shell thickness (and more generally, the velocity profile).

The number of photons emitted from the shell at a given time is proportional to the continuum flux from the supernova photosphere. 
Therefore, the form of the light curve between $\sim 2765 - 2834$ \AA \ (for an expansion velocity $\sim 3300 \kms$) is important. We determine the continuum in this range from the available spectra, shown as black points in Supplementary~Fig.~1. Unfortunately, we lack spectra on the rise of the light curve, and have to make assumptions about the flux in this wavelength range prior to the first spectrum.

As a parametrization we use a generalized Gaussian distribution given by
 \begin{equation}
 F_{\lambda}(2800 \text{\AA}) = C \frac{e^{-y^2/2}}{\alpha -\kappa( t_{\rm day}-\zeta)}
 \label{eq1}
\end{equation}
where $y=-\ln[1-\kappa (t_{\rm day}-\zeta)/\alpha]/\kappa$. As our standard model we use $\zeta=41$ days, $\alpha=23$ days and the shape parameter $\kappa=-0.28$. This gives a good fit to the continuum light curve at 2800 \AA \ in Supplementary~Fig.~1 after the peak, and also fit reasonable assumptions about the 2800 \AA flux on the rise, as we describe below.

For the light curve before the peak we have to refer to more indirect arguments. The 2800 \AA \ region falls between the u and g band for iPTF16eh. While the g-band has a good coverage also before the peak, we have no observations in the u band for iPTF16eh. As a guide
we therefore use the observations and modeling for PTF12dam\cite{nsj+13,tnb+17}, which as
discussed earlier has similar light curve and spectra. For PTF12dam
2800 \AA \ falls between the uvw1 ($\sim 2350$\AA) and u ($\sim 3200$\AA) bands. Noting that the g-band for
iPTF16eh corresponds closely to the u-band for PTF12dam in redshift,
we find that the u-band model (Fig. 7 in\cite{tnb+17}) provides a
good fit to the g-band light curve of iPTF16eh, both before and after
the peak, using a modest stretch factor of 1.15 compared to the
PTF12dam model. In addition, the PTF12dam uvw1-band model gives a good
fit to the observed 2800 \AA \ observations of iPTF16eh after the peak,
now with a stretch factor of 0.9. With these stretch factors both the uvw1- and u-band models result in very similar
light curves before the peak, while after the peak the uvw1-band light
curve drops much faster than the u-band because of stronger
blanketing. This can also be seen for iPTF16eh by comparing the 2800 \AA \
and the g-band light curves in Supplementary~Fig.~1. The physical motivation for the stretch factors is that the diffusion time scale depends on the combination $\kappa M/V$, where $\kappa$ is the opacity, $M$ is the ejecta mass and $V$ the expansion velocity. All these can differ between the two supernovae, in particular the opacity, which is also strongly wavelength dependent. The diffusion form of the light curve should, however, be more robust.
Summarizing, Eq. (\ref{eq1}) should give a good
representation of the 2800 \AA \ light curve both before and after the
peak with the above parameters.
However, because of the uncertainty in the flux at 2800 \AA \ before the peak we discuss departures from Eq. (\ref{eq1}) below.

Our model photosphere emits in random directions in the interval 2760 - 2840 \AA. The doublet nature of the lines complicates the radiative transfer, since a photon scattered in the $\wl 2795.5$ line may be further scattered by the $\wl 2802.7$ line. 
Only photons below $2802.7$ \ \AA \ will be scattered, while those above will escape directly or be absorbed by the photosphere. The time for each photon from its emission until it arrives at a common surface relative to the line of sight is calculated. 
For a single line and homologous expansion a photon emitted at $t_{\rm emiss}$ from a radius $r_1$ at an angle with cosine $\mu_1$ will arrive at 
 \begin{equation}
    t = t_{\rm emiss} +  (R_{\rm out}  x - r_2 \mu_2)/c
\end{equation}
Here $\mu_2$ is the cosine of the angle relative to the radial direction of the scattered photon at the radius of scattering $r_2$, given by 
 \begin{equation}
r_2 = [r_1^2 + (x R_{\rm out})^2 + 2 x R_{\rm out} r_1 \mu_1]^{1/2}
\label{eq_r2}
\end{equation}
and the dimensionless frequency shift is 
\begin{equation}
x = \frac{\nu - \nu_0}{\nu_0} \frac{c}{V_{\rm max}} .
\end{equation}
Equation (\ref{eq_r2}) describes a spherical surface with radius $x R_{\rm out}$ centered at $r_1$.
For the constant velocity case similar equations are obtained, although somewhat more complicated.  We assume isotropic scattering in the rest frame of the gas. 

The thickness of the shell influences both the shape of the line profile and the flux relative to the nearby continuum, being roughly proportional to the cross sectional area of the shell, $\pi(R_{\rm out}^2-R_{\rm in}^2)$ (Supplementary~Fig.~5). The fraction scattered is then $\sim \Delta R/R$, where $\Delta R$ is the thickness of the shell. From the observed ratio of the luminosity of the continuum emission at 2800 \AA \ and the luminosity of the emission at the late epochs we find that the range $R_{\rm in}= 130$ light days and $R_{\rm out}=137$ light days gives a good agreement with the observations. This is only weakly dependent on the velocity law used. 

If it had not been for the importance of the large light travel time compared to the evolutionary time scale of the photospheric emission this would have resulted in an ordinary P-Cygni line, with nearly zero equivalent width.

For the resonance scattering scenario to work the Mg~II lines have to be optically thick. Assuming a homologous expansion and using the derived parameters for the volume of the shell from the main section, the optical depth of the Mg~II line can be estimated as 
\begin{equation} 
\tau  = \frac{A_{21}  \lambda^3 n_1 t}{8 \pi g_1} \approx 3.2 \times 10^{6} X({\rm Mg II}) \left(\frac{M_{\rm shell}}{\Msun}\right) \ ,
\label{eq_tau}
\end{equation}
where $A_{21}$ is the transition rate, $\lambda$ the wavelength, $g_1$ the statistical weight of the lower state (equal to 2), $n_1$ the number density in the ground state and    $X($Mg~II$)$ the Mg~II ionic abundance. The time  $t$ is taken as the time since ejection discussed earlier, $\sim 35$ years. The solar Mg abundance is $\sim 4\times 10^{-5}$  by number. For a shell with a mass $\gtrsim 1 \Msun$ (as we argue in the main text) the optical depth can therefore be large even if the abundance is less than solar, as long as most of the Mg is in Mg~II. This is likely to be the case, unless there is a very strong flux above $\sim 15$ eV. 

In the main section of the paper we show the evolution of the line profiles for the 'standard' model with $R_{\rm in}=130$ light days and $R_{\rm out}=137$ light days and the parametrization of Eq. (\ref{eq1}) of the continuum flux. To illustrate the effects of the doublet scattering we show in the upper right panel in Supplementary~Fig.~4 a model with the same parameters, but for only a single line at 2800 \AA.   
Compared to the single line simulation one notes the double peak nature of the line profiles. Because of the scattering from the blue to the red component the line profile is, however, not a simple sum of the two components. The roughly double intensity of the line for the doublet is a result of the twice as large continuum covering by the doublet. 

The influence of the unobserved continuum light curve before the peak is also shown in Supplementary~Fig.~4 for a model where we have assumed a constant flux before the peak and after that the observed decline. As for the case of the burst in the main section, the line profile reflects this as a plateau on the long wavelength side of the line. Finally we show the effect of different shell parameters, $R_{\rm in}=150$ light days, $R_{\rm out}=160$ light days for the standard light curve. Compared to the standard case this results in narrower line profiles, but, as expected, a slower velocity evolution.

The fading part of the Mg II flux gives a precise estimate of the outer radius of the shell, given by $R_{\rm out}=c(t_{\rm fade}-\tau)/2$, where $t_{\rm fade} \approx 375$ days is the time when the line is too weak to be seen.  $\tau$ is the effective length of the light curve, which we define as the time since explosion corresponding to 95\% of the integrated luminosity at 2800 \AA. Our model for the light curve (Eq. \ref{eq1}) gives $\tau \approx 90$ days. From simulations of the light curve of the Mg II line for different shell parameters, we estimate the error in the outer radius to be $\pm 7$ light days, and a similar number for the inner radius. The thickness should, however, be $\sim 5\%$ to give the correct absolute line fluxes.

\subsection{Ruling out alternative explanations of the narrow Mg~II lines}
We can rule out that the Mg~II emission lines are originating in the SN ejecta due to a peculiar chemical composition, since this would require a Mg abundance close to 100 \%. Nuclear burning zones with a high abundance of Mg nearly always have even higher abundance of O and Ne. One would therefore expect strong emission lines of O~I-III, as well as Ne~III-V and Mg~I from the shell.  

Direct interaction with a CS shell, as in SN 2001em\cite{cc06} and SN 2014C\cite{mmk+15,mkm+17,ahm+17,bkm+17}, can be ruled out from the time scale and line shifts, as well as the absence of He and metal lines, which would dominate the cooling for abundances typical of H-, He- or O-rich zones. A further argument is based on the observed line shifts with time. For an expansion velocity $V_{\rm exp}$ the shell has to be at $t_{\rm shell} V_{\rm exp}$, where $t_{\rm shell}$ is the first detection of the lines. We adopt a conservative value of $t_{\rm shell} \sim 100$ days. Even for an extreme velocity of $0.2 c$ the shell would have to be at $\lesssim 5 \times 10^{16}$ cm. Light travelling time effects would then be only of order $V_{\rm exp}/c$, and would not be important for explaining the drifting velocity. A decreasing velocity of the shock from negative to positive would also be extremely difficult to understand. 

Based on the shell velocity, an echo from the wind of a Wolf-Rayet star could be a possibility. This can, however, be refuted on several grounds. First, the state of ionization of a WR wind is considerable higher, with typical ions being C~IV, N~V and O~VI, but with no absorption lines from Mg~II. If it was instead coming from the interaction region between the WR wind and the slow material from a previous RSG or LBV stage the velocity would be of the order of hundreds of $\kms$ and not $\sim 3300 \kms$.

\subsection{Constraints on the shell mass and composition}

The Mg~II doublet is one of the few strong resonance lines of the abundant elements  in the observed part of the spectrum. The other possible resonance lines are the Ca~II $\wll 3968.5, 4226.7$ H and K lines, the Na~I D $\wll 5890.0, 5895.9$ doublet and the K~I  $\wll 7664.9, 7699.0$ doublet. In addition, there are weaker lines from Fe~II and Ti~II. 

Both Na~I and K~I are easily ionized out to very large distances by the radiation from the SN due to their low ionization potentials 5.14 and 4.34 eV, respectively. With regard to the Ca~II lines the solar Ca abundance is only $\sim 5 \%$ of the Mg abundance, while the atomic line parameters, like the transition rate and excitation energy, are comparable. In addition, the ionization potential of Ca~II is only 11.82 eV, below the H~I and O~I ionization edges, and therefore subject to a strong ionizing flux from the SN. Mg~II with an ionization potential of 15.03 eV is more resistant to this. Given the hot SN continuum, it is therefore not surprising that no Ca~II absorption lines are seen from the shell.

If the progenitor had some of the H envelope left before the first ejection this would temper the velocity and also change the time between the first pulse and core collapse. Unfortunately, we do not have any strong constraints on the composition of the shell from observations. An upper limit to the H mass can be obtained from the H$\alpha$ luminosity, $L _{H{\alpha}}$. Assuming the shell is H-dominated, and given the high radiation temperature at the first observations (Supplementary~Fig.~2) also fully ionized, we calculate

\begin{equation} 
L _{H{\alpha}} 
\approx 3.5 \times 10^{36} f^{-1} \left(\frac{T_{\rm e}}{10^4 {\rm K}}\right)^{-0.94} \left(\frac{M_{\rm shell}}{\Msun}\right)^2 
\left(\frac{R_{\rm shell}}{5\times 10^{17} \rm cm}\right)^{-3} \left(\frac{\Delta R_{\rm shell}/R_{\rm shell}}{0.1}\right)^{-1} \ \ergs ,
\end{equation}
where 
$f$ is the filling factor of the shell. For $R_{\rm shell} \approx 3.5 \times 10^{17}$ cm and $\Delta R_{\rm shell}/R_{\rm shell} \approx 0.06$ one obtains $L _{H{\alpha}}\approx 9.6 \times 10^{36} (M_{\rm shell}/\Msun) f^{-1} \ergs$. To estimate an upper limit to the H$\alpha$ flux we use the February 2017 spectrum, which has the best S/N of the spectra where the peak of the line is well separated from the H$\alpha$ from the host galaxy (using the velocity shift from the Mg~II line). Using also the same line profile as for the Mg II line at this epoch, we find an upper limit of $F_{H\alpha}\lesssim 1.2 \times 10^{-17} \ergs {\rm cm}^{-2}$, corresponding to  $L_{H\alpha} \lesssim 8.1\times 10^{39} \ergs$ for standard cosmological parameters.  Any emission from a hydrogen dominated shell with $M \lesssim 28 f \Msun$ would therefore be difficult to detect, unless the filling factor is small. This is mainly a result of the large distance to the shell and therefore low density. 

\subsection{Geometry of the shell}
As argued in the main text, the fact that the shell is seen in both emission and absorption at very similar velocities, as well as the smooth velocity evolution (Fig.~\ref{fig:mgii_prop}) argues for the shell being symmetric. Further information on the geometry comes from comparing the absorbed and re-emitted flux, which constrains the covering fraction.

From the scattered emission measured by the time integrated absorbed flux in the two doublet components at 2767 and 2775 \AA \ one can estimate the total number of of photons scattered by the shell. For a spherical shell with a complete covering factor this should be equal to the total number of emitted photons in the emission component, ignoring the (negligible) absorption by the photosphere. To estimate the total  scattered flux we have measured the equivalent width of the blue component, which is not affected by the scattered emission. We determine the continuum level from the average flux on the blue side of the lines between 2750 to 2763 \AA. From the spectra with the highest signal to noise in the continuum (days 52, 28, 82, 100, 125) we find an equivalent width of 2.0 \AA \ in the rest frame of the SN. Using the continuum light curve in Eq. (\ref{eq1}) with $C=4.5 \times 10^{-15} \ \ergs {\rm cm}^{-2} \ \text{\AA}^{-1}$ and multiplying this with $2.0$ \AA \ we can estimate the total absorbed flux  in the blue  2795.5 \AA \ component to $\sim 1.8 \times 10^{-9} \ {\rm erg \ cm^{-2}}$. Assuming that the red 2802.7 \AA \ component contributes the same we arrive at $\sim 3.6 \times 10^{-9} \ {\rm erg \ cm^{-2}}$. 

To estimate the total flux in the emission component we assume from Fig.~\ref{fig:mgii_prop} a constant flux of $\sim 1.75 \times 10^{-16} \ {\rm erg \ cm^{-2} \ s^{-1}}$ during 250 days, taking the rise and decline of the flux into account. We then find a total fluence of $\sim 3.8 \times 10^{-9} \ {\rm erg \ cm^{-2}}$ for the emission component. Given the uncertainties in both the absorbed flux (lack of spectra in the rising part of the light curve and the general continuum level for the equivalent width estimate) and also in the the emitted flux (including the lack of observations in the middle part of the light curve), these two values are consistent within the errors. For an incomplete covering factor, e.g. a ring or torus, one would expect the emission flux to be much less than the scattered flux in the absorptions. 

Although the covering factor should be large, the shell may in principle be non-spherical, e.g., bipolar, as in Eta Carinae, or simply irregular. The evolution of both the peak velocity and flux, however, put constraints also on the shape of the shell.In particular, the very close to linear evolution of the peak velocity with time (Fig. \ref{fig:mgii_prop}) is very difficult to get for an irregular shell, and requires either a spherical shell or a bipolar geometry with a special orientation.

Although extending the simulations we performed for a spherical shell
to more general geometries is beyond of scope of this paper, one can see some general trends by examining the intersections of the echo parabolas, similar to that of Supplementary~Fig.~5, with shells of more complicated geometry. This would correspond to the velocity evolution for a short supernova flash emitted 50 days after explosion by a thin shell, and should therefore not be directly compared to our Monte Carlo simulations, where the echo from the whole supernova light curve is taken into account. The delay of 50 days corresponds approximately to the middle of the light curve. The main effects of different geometries should, however, be insensitive to this. 
 
 As examples of this we have examined ellipsoids with two different, limiting orientations relative to the line of sight. The simplest cases are for oblate and prolate spheroids oriented along the line of sight, which in this case have cylindrical symmetry along the line of sight. In the upper panel of Supplementary~Fig.~6 we show the shape of these for an axial ratio of 1.37 between the major and  minor axes. The extent along the line of sight is set by the duration of the echo and is taken to be 137 light days in all cases. The minor (major) axis is therefore 100 and 188 light days, respectively. The evolution of the peak velocity is also sensitive to the assumptions about the velocity of the shell as function of radius, so a complete coverage of all possibilities is difficult. Here we examine the cases of homologeous expansion, $V(r) \propto r$, and constant velocity for the shell, shown in the lower panels of Supplementary~Fig.~6, together with the observed evolution. As can be seen from this figure, even a modest axial ratio of 1.37 has an appreciable effect of the velocity evolution, and it is clear that for these models the simple spherical model is also the most accurate. Although we have not calculated detailed light curves, also the flux evolution should be affected, since the projected area of the shell within the light parabola will be different depending on the geometry.

\subsection{Circumstellar emission in other related SNe}
While iPTF16eh is the first supernova to show a resonance echo, there are several examples of supernovae where the presence of late-time mass loss is inferred. For example, the SLSN-I iPTF13ehe\cite{yqo+15}, iPTF15esb and iPTF16bad\cite{ylp+17} all show broad H$\alpha$ emission at late epochs, interpreted as emission from a shocked shell excited by the expanding ejecta. The inferred radius for the shells in these cases is a few times $10^{16}$~cm, 
an order of magnitude smaller
than the shell radius inferred for iPTF16eh.

The $\sim 4000 \kms$ width of the lines in iPTF13ehe were by \cite{yqo+15} interpreted as the thermal velocity of hydrogen, while the expansion velocity was only $\sim 300 \kms$. This would correspond to a temperature of $\gtrsim 10^8$ K, at what temperature hydrogen is a very inefficient emitter. Alternatively, the width may be interpreted as the expansion velocity of the shocked shell behind a radiative shock. With an ejecta velocity $\sim 13,000 \kms$ the shell density should be $\sim (13,000/4000)^2 \approx 10$ times that of the ejecta. A radiative shock would, however, require a very high density $\gtrsim 10^9 (V_{\rm s}/4000 \kms) \ccm$ of the pre-shock shell, and a factor $\sim 10^4 - 10^5$ higher in the cool H$\alpha$ emitting gas.
An alternative could be that the excitation is instead coming from ionizing radiation originating in the supernova explosion. This could either be from an early burst connected to the shock breakout or from the later emission from the photosphere if it is hot enough. The observed H$\alpha$ would then come from the recombination of the ionized shell. Except for the much higher velocity, this is similar to the mechanism which excited the ring of SN 1987A\cite{lf96}. This has the important implication that the distance to the H$\alpha$ emitting material could be much larger, and that the velocity of the H$\alpha$ line would be characteristic of the shell, in the same way as discussed here for iPTF16eh. We also note that the line profiles of both iPTF13ehe and iPTF15bad have 'shoulders', more similar to the boxy profiles of an expanding shell than Gaussian profiles of a thermal gas. For the same geometric parameters as for iPTF16eh this would require a mass $\gtrsim{ 10} \Msun$ for a filling factor $\gtrsim 0.1$. Because of the unknown composition of the iPTF16eh shell and the uncertainty in the excitation mechanism of the other SNe, reflected in the distance to their shells, it is difficult to judge how similar these cases are.

There are other SNe showing dense shells of CS gas. The condition to observe an echo or not is that $\tau \ll R_{\rm shell}/c$  where $\tau$ is the characteristic duration of the exciting radiation. This was the case of SN 1987A, where the exciting radiation of the circumstellar ring was the shock breakout radiation which had a time scale of minutes, while the ring is at a distance of $\sim 200$ light days, and an echo was clearly observed. In this case the emission from the ring was seen mainly as UV lines\cite{fcg+89}. Because the expansion velocity of the ring was only $\sim 10 \kms$, only a marginal shift in wavelength with time was seen. Moreover, the recombination time of the gas was long, which smoothed the evolution\cite{lf96}. Except for the N V $\wll 1238, 1242$ lines, the excitation was mainly by collisions in the hot gas. 

Two other related cases are the stripped SNe 2001em\cite{cc06} and SN 2014C\cite{mmk+15,mkm+17,ahm+17,bkm+17}. For the latter, radio, optical, and X-ray observations have revealed a shell of hydrogen rich gas at $\sim 6 \times 10^{16}$ cm with a mass of $\sim 1 \Msun$. The analysis of SN 2001em yielded a similar radius for the shell, with a somewhat higher mass, $2 - 3 \Msun$\cite{cc06}. The light travel time across these shells $\sim 20$ days, which is shorter than the width of the light curve. Echoes, such as the ones we infer for iPTF16eh, are therefore not likely to be important for these.

\subsection{Constraints on the progenitor}
As discussed in the main text, we believe that the most likely explanation for the shell is a PPI ejection before the final explosion, although an LBV related eruption, similar to that of Eta Carinae, cannot be excluded. However, as discussed in Ref.~22\nocite{woo17}, an LBV may also be a result of PPI ejections. Here, we compare the inferred properties of iPTF16eh more closely to PPI models, to see what constraints can be obtained for the progenitor star.

The observations give a shell velocity, $\sim 3300 \kms$, with a radius, $\sim 3.5 \times 10^{17}$~cm, implying a time scale of $\sim 32$ years, if undecelerated. From the limit of the H$\alpha$ luminosity we derived a weak limit on the hydrogen mass of $M \lesssim 28 f \Msun$. It could therefore be dominated by either hydrogen or helium. The shell is likely to be close to spherical with a large covering factor.

For the PPISNe the main parameters for the time between the first shell ejection and the subsequent ones and the final core collapse is the mass of the He core. This time increases rapidly above $\sim 50 \Msun$ (Table 1 in Ref.~22\nocite{woo17}). For a bare He core without rotation a time scale of $\sim 32$ years corresponds to a He core mass of $\sim 54 \Msun$.

An example of a model with a low mass H envelope that gives about the right shell ejection properties is B115-5 in Table 4 of Ref.~22\nocite{woo17}. This was (by construction) a blue supergiant (or LBV, whose properties would have been similar) with a hydrogen envelope mass of $4.9 \Msun$ (though the hydrogen mass fraction in this envelope was only 0.2). The first two pulses ejected $6.7 \Msun$ including all this envelope. The total kinetic energy for that ejection was $6.6 \times 10^{50}$ erg.  5.2 years later two additional pulses separated by about 20 days ejected an additional $4.0 \Msun$ with a total kinetic energy of $3.5\times 10^{50}$ erg. This matter was mostly He with some new C, O, Ne, and Mg. About 10 days later the remaining $45.9 \Msun$ core collapsed, probably to a black hole.

At the time of the final outburst the shell from the first pair of pulses was moving at $3100 \kms$ at a radius of about $5 \times 10^{16}$ cm. This is a similar shell velocity but too small of a shell radius compared to iPTF16eh. 
The delay time is, however, very sensitive to the mass of the core and to some extent the mass of the H envelope. A $20 \Msun$ envelope on the same model gave a delay of 14 years, but also a lower shell velocity. While none of the models in Ref.~22 are precise matches to iPTF16eh, these trends suggest that to get a fast-moving shell at the distance of iPTF16eh (implying a delay of $\sim 30$ years), likely a combination of a somewhat heavier star than model B115-5, with a hydrogen envelope mass of $\sim 5 \Msun$, is needed.

In summary, a PPISN can eject several solar masses of material years to
decades before the final explosion. For a 30 year delay, the main
sequence mass is close to $115 \Msun$ and the He core mass close to $52 \Msun$. Given the severe energy restrictions on the PPISN model 
the mass of the fossil shell, possibly H-rich, can be no more than about $10 \Msun$.

\subsection{Powering the supernova light curve}

While the above models agree well with the velocity and time scale of the observed shell, the most problematic aspect is that the SNe predicted for these models are not superluminous.  The light curve of the final explosion for model B115-5, shown in Fig.~22b of Ref.~22, lacks the duration and high luminosity of iPTF16eh. In general, it is hard to get it much brighter without more mass,  meaning lower ejection speeds, in a pure PPISN model. Further, we find a total radiated energy of $\sim 1.3 \times 10^{51}$ erg for iPTF16eh, while the maximum energy of the pure PPISN models in Ref.~22\nocite{woo17} is $\sim 5 \times 10^{50}$ erg.
Additional energy input may therefore be needed and there are several possibilities for this, all, however, somewhat speculative.

A pure magnetar model 
could provide the extra energy. This would correspond to the 'standard' magnetar model for a SLSN-I\cite{kb10}. There is, however, for these no natural explanation for the shell formation. An LBV eruption connected to the last burning stages may be a possibility\cite{qs12}, although the cause is not well understood.

A second possibility is that a high mass PPISN also makes a magnetar. The problem is here the high mass of the Fe core for these massive stars, which tends to result in a collapse to a black hole rather than a neutron star. A possible scenario is, however, that the PPISN first ejects the observed shell and at the core collapse a milli-second magnetar forms from the 2.1 $\Msun$ Fe core. The magnetar is born about 10 days after the last shell ejection. Initially the field has to be very strong ($\gtrsim 10^{15}$ G) injecting $\sim 10^{52}$ erg to blow up the star. To power the light curve over a longer period the magnetic field has to decay to  $\lesssim 10^{14}$ G before the neutron star radiates all its energy. This obviously needs a high degree of fine tuning.

In the final scenario a black hole of about 40 $\Msun$ is born, but matter at its outer boundary has enough rotation and magnetic field to form a disk and make jets. This takes more rotation than the  above options and may be difficult to achieve if there has been a lot of mass loss or the star has been a giant. But if a disk forms $1 \Msun$ accreted mass could give $\sim 10^{53}$ erg, which might come out as a broad jet. Interaction with matter at $10^{14}-10^{15}$ cm from the mass ejections a week before may then lead to a bright enough light curve. The event would, however, probably be asymmetric. 

We note that both the magnetar and the accreting black hole scenario will require a rapidly rotating progenitor star. This may be in conflict with our constraints on the shell geometry, which is best fit by a spherical shell.

\subsection{Data Availability Statement}
The photometry of iPTF16eh is available in Supplementary Table~1, and the spectra are available from WISeREP\cite{yg12} ($\rm{http://wiserep.weizmann.ac.il/}$). In general, the data that support the plots within this paper and other findings of this study are available from the corresponding author upon reasonable request.

\subsection{Code Availability Statement}
The light echo modeling code is available from Claes Fransson (claes@astro.su.se) upon request.

\end{methods}


\setcounter{enumiv}{\value{firstbib}}


\clearpage

\begin{addendum}

 \item[Supplementary Information]

\begin{SIfigure}
\centering
\includegraphics[width=6in]{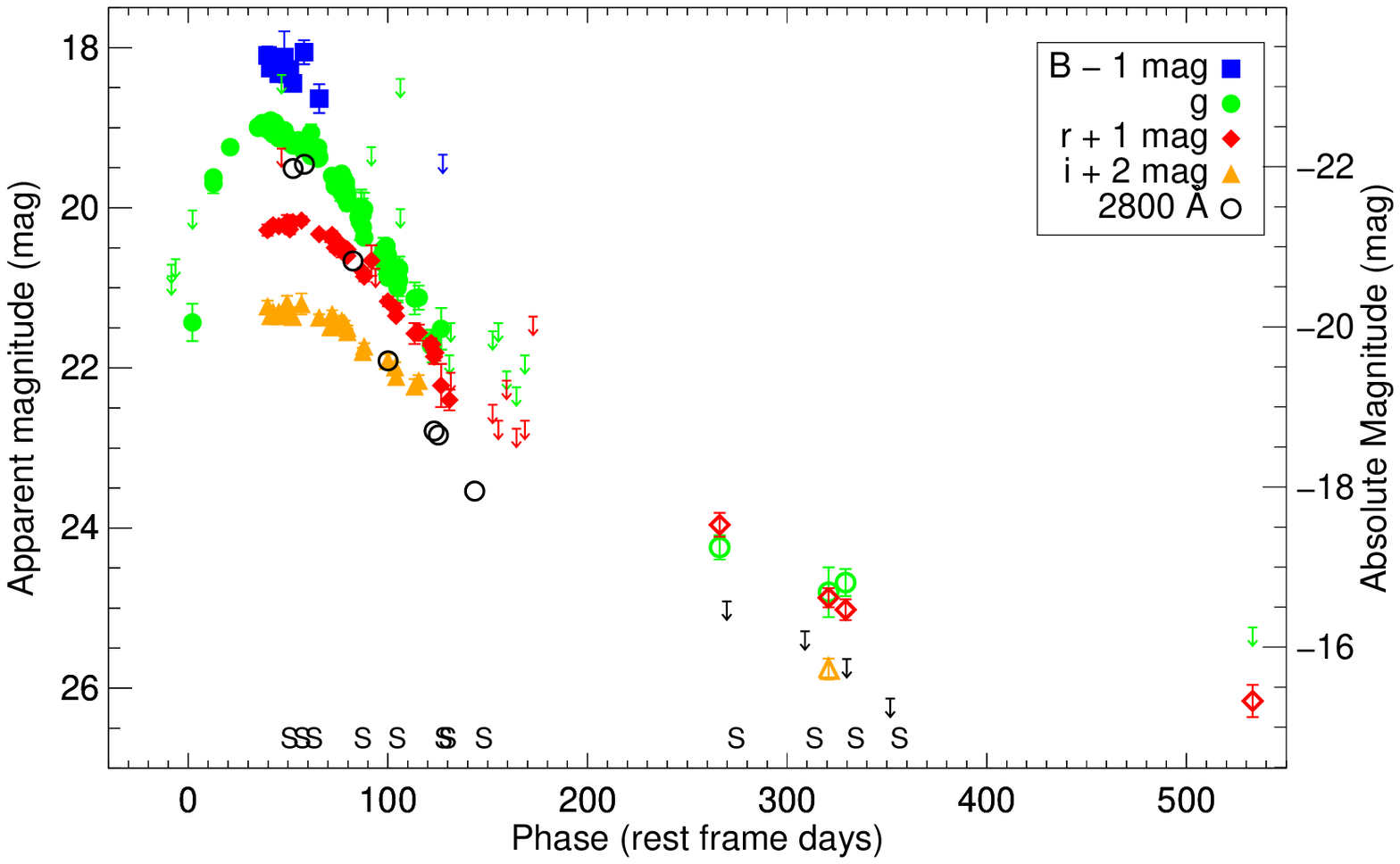}
\caption{Observed light curve of iPTF16eh. Data is mainly from the P48 and P60 telescopes, and is listed in Supplementary~Table~\ref{tab:phot}. Filters have been offset for clarity as indicated in the legend, and dates of the spectroscopy epochs are marked with ‘S’ symbols along the bottom axis. Upper limits are 5$\sigma$. The open symbols after 200~days show unsubtracted photometry, and may have a significant host galaxy contribution. Open black circles show the flux at a rest-frame wavelength of 2800~\AA\,, as measured from the continuum flux in the spectra in the region 2765 -- 2834~\AA.
\label{fig:obslc}}
\end{SIfigure}

\newpage

\begin{SIfigure}
    \centering
    \includegraphics{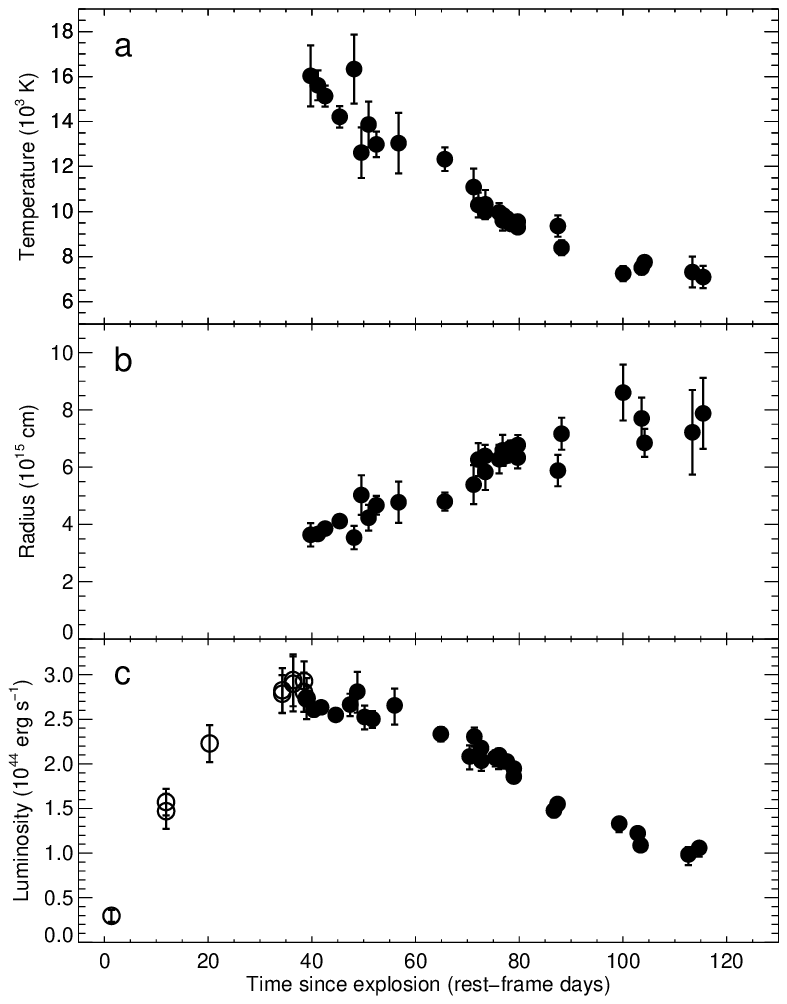}
    \caption{Evolution of (a) the blackbody temperature, (b) the blackbody radius, and (c) the pseudo-bolometric luminosity. Blackbody parameters are derived by fitting a Planck function to the observed photometry using IDL's {\tt mpfit} routine, and the error bars plotted are the 1$\sigma$ uncertainties given the observed photometry. The pseudo-bolometric luminosity is calculated by integrating the observed flux as well as a blackbody tail redwards of the observed bands, and the plotted uncertainties include both these components. As we do not have multiband data on the rise of the light curve, the points on the rise are calculated scaling from the $g$-band light curve and assuming a constant bolometric correction (open circles). Any extra (systematic) uncertainty from this assumption is not included in the error bars, which reflect the statistical uncertainties only.
    \label{fig:bb}}
\end{SIfigure}

\newpage

\begin{SIfigure}
    \centering
    \includegraphics{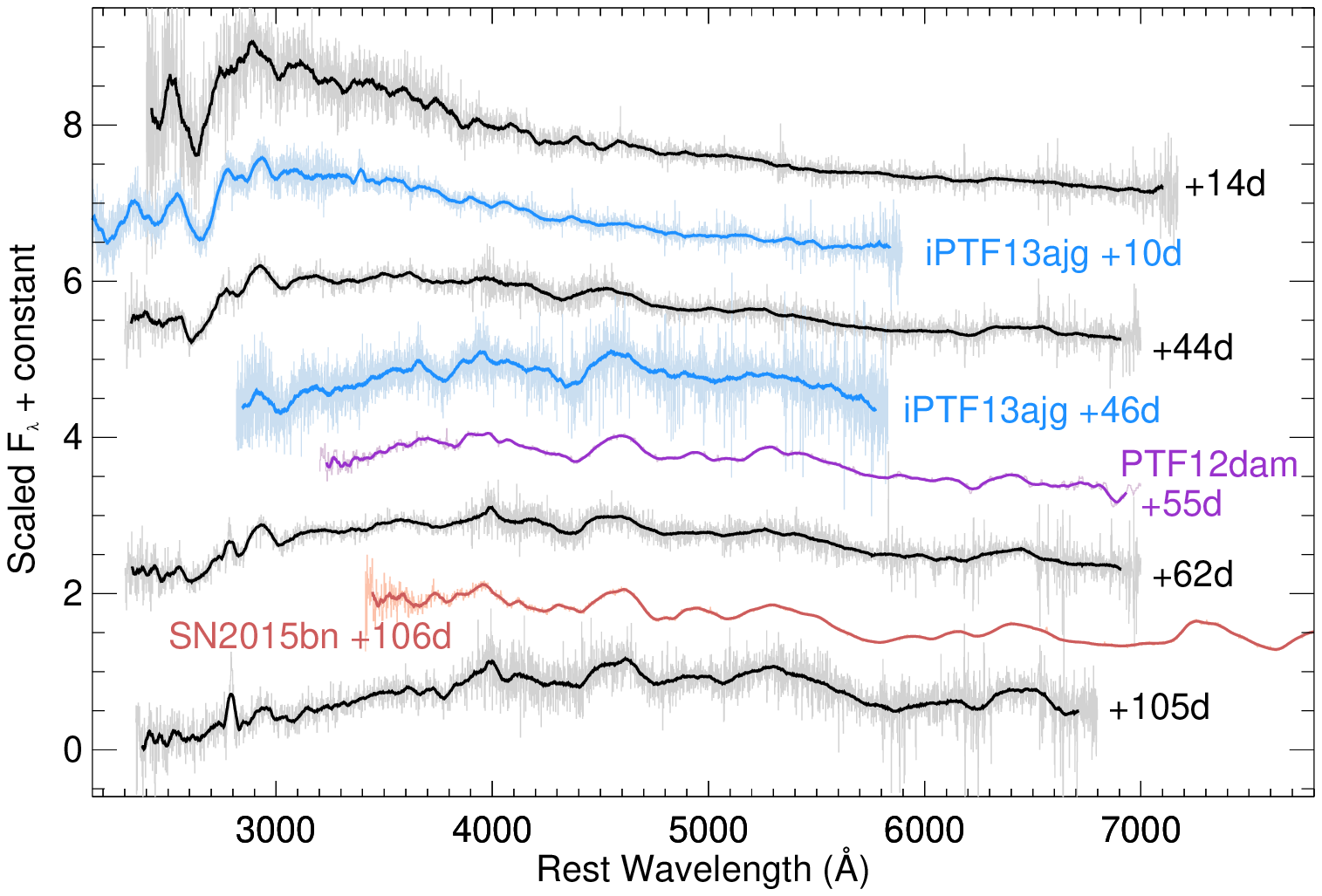}
    \caption{Spectra of iPTF16eh (black) compared to the well-studied SLSNe PTF12dam\cite{nsj+13}, iPTF13ajg\cite{vsg+14}, and SN\,2015bn\cite{nbs+16}. To facilitate comparisons, supernova phases are labeled relative to peak light (MJD 57424.3 in the case of iPTF16eh). The spectra are smoothed by a Savitzky-Golay filter, with the unsmoothed data shown in the background. The spectra have been normalized to the flux level at 4000~\AA\, and are offset from each other by one scale unit. 
    \label{fig:spec_comp}}
\end{SIfigure}

\newpage
\begin{SIfigure}
\centering
\includegraphics[width=6.in]{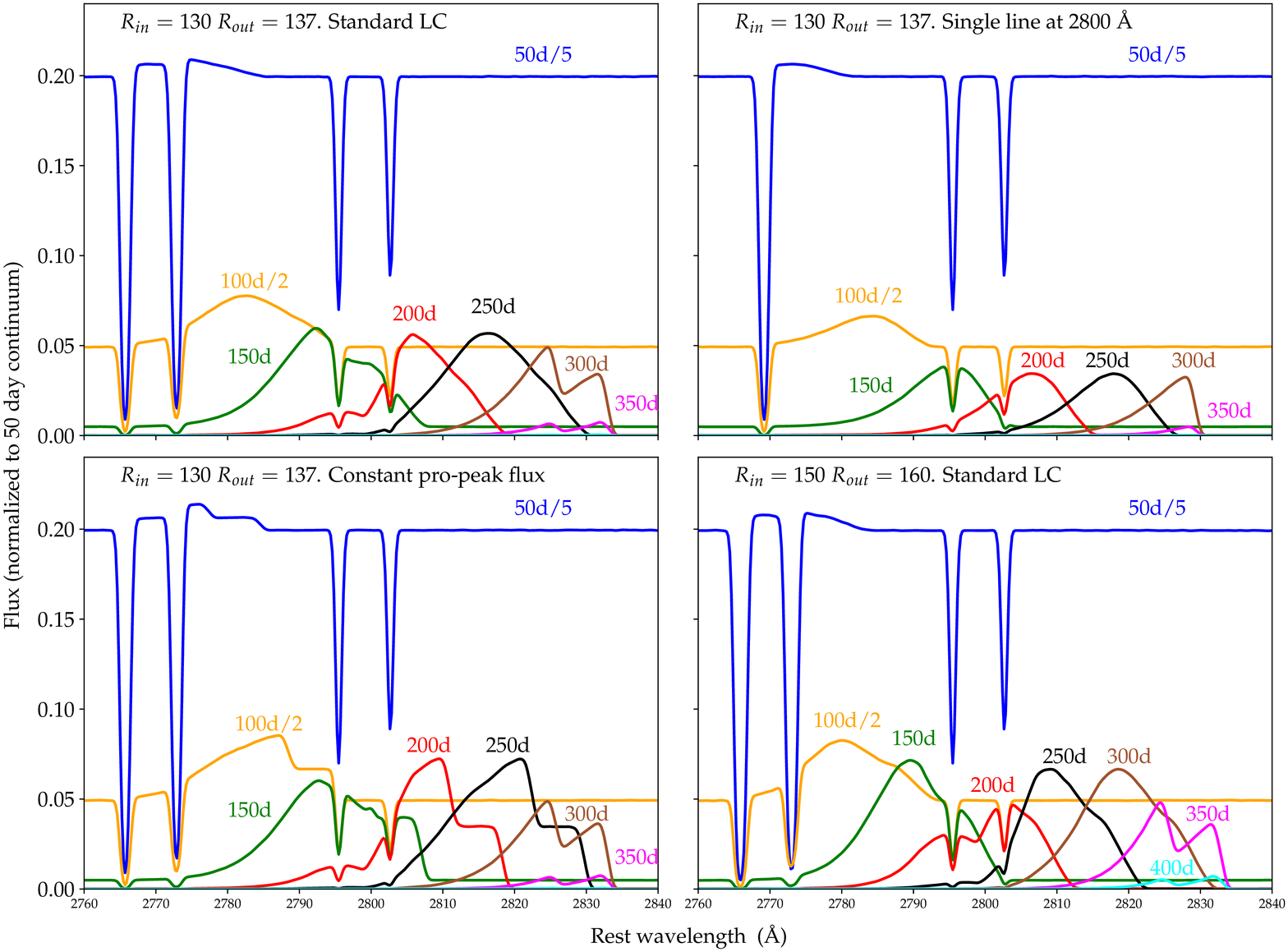}
\caption{Time sequence of simulated spectra in the Mg~II wavelength region, similar to Figure~3 of the main article, for different choices of shell parameters and input light curves. \textit{Upper left:} Standard values for the light curve and shell, $R_{\rm in}=130$ light days, $R_{\rm in}=137$ light days. \textit{Upper right:} Same parameters, but with only a single line at a rest wavelength 2800 \AA.  Note the double-peaked line profiles, the broader lines, as well as the nearly double intensity in the doublet case.
\textit{Lower left:} Same shell parameters, but constant continuum flux before the peak. Note the flat red plateau in the line profile. \textit{Lower right:} Effect of changing the parameters of the shell to $R_{\rm in}=150$ light days, $R_{\rm in}=160$ light days. Standard light curve. Note the slower evolution in wavelength and narrower line profiles. No smoothing due to the instrumental resolution has been applied too these simulations. 
\label{fig:single_double}}
\end{SIfigure}

\clearpage

\begin{SIfigure}
\centering
\includegraphics[width=6.in]{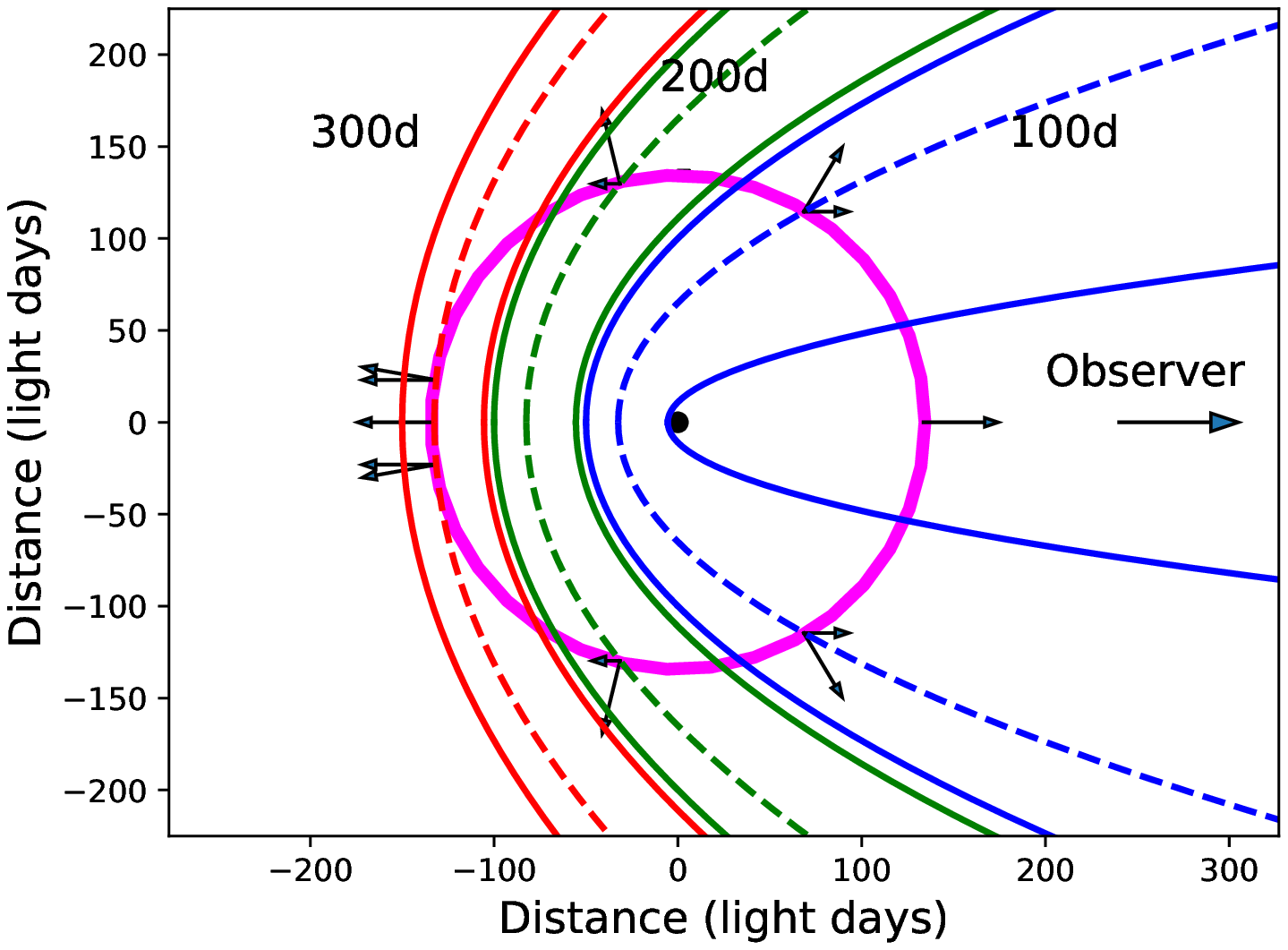}
\caption{Approximate geometry of the shell and light echo parabolas, corresponding to three different observed epochs, 100, 200 and 300 days after explosion. The outer parabola corresponds to the shock breakout and the dashed to the peak luminosity, assumed to be at $\sim 35$ days after shock breakout. The inner parabola, representing the end of the light curve, is defined by the time after shock breakout corresponding to 95\% of the integrated luminosity of the model light curve, here 89 days. The radial and horizontal arrows correspond to the expansion velocity of the ring, $\sim 3300 \kms$, and the line of sight velocity for the peak of the light curve, respectively. 
\label{fig:echo_parabola}}
\end{SIfigure}

\begin{SIfigure}
\centering
\includegraphics[width=4.in]{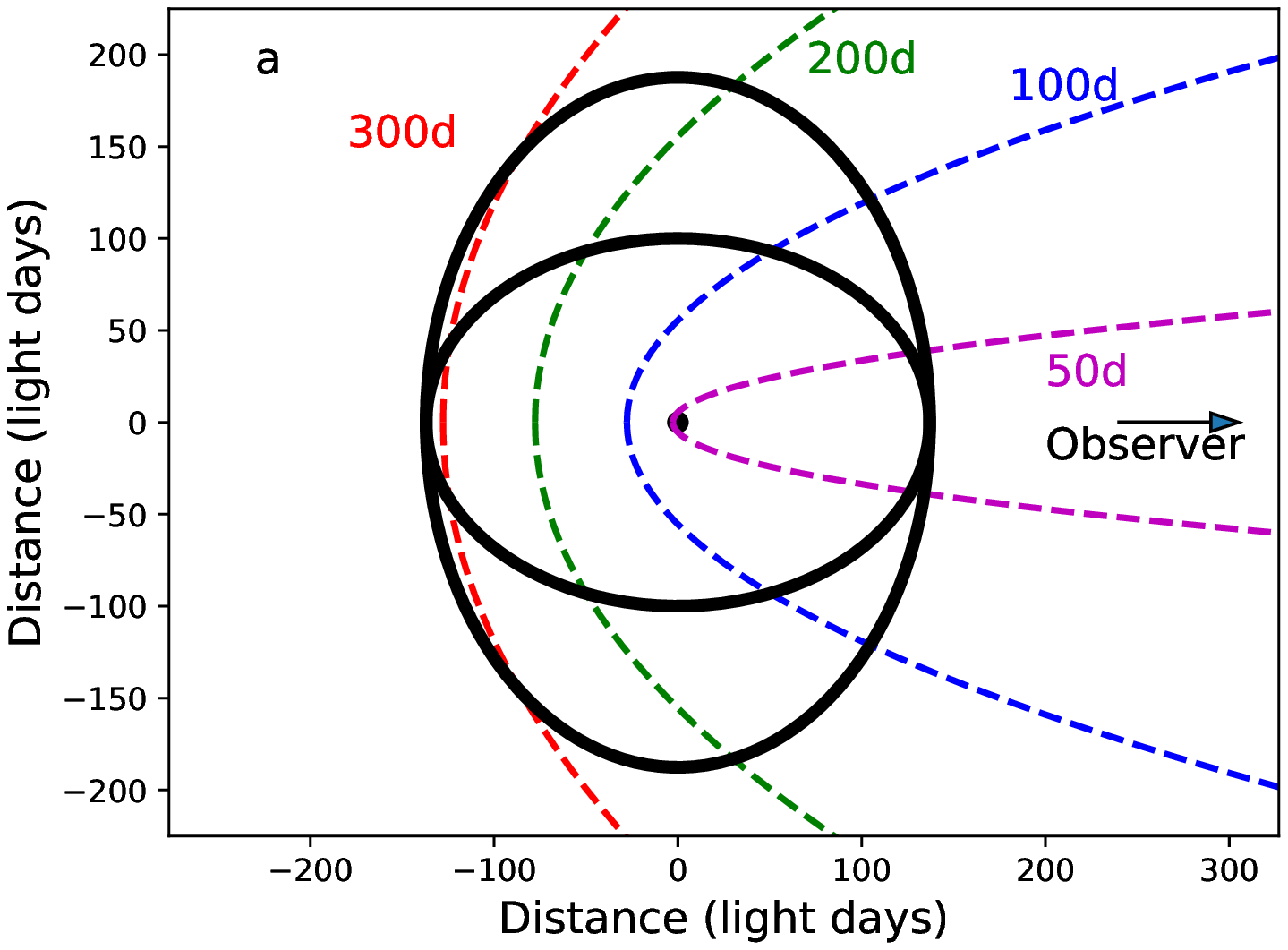}
\includegraphics[width=3.in]{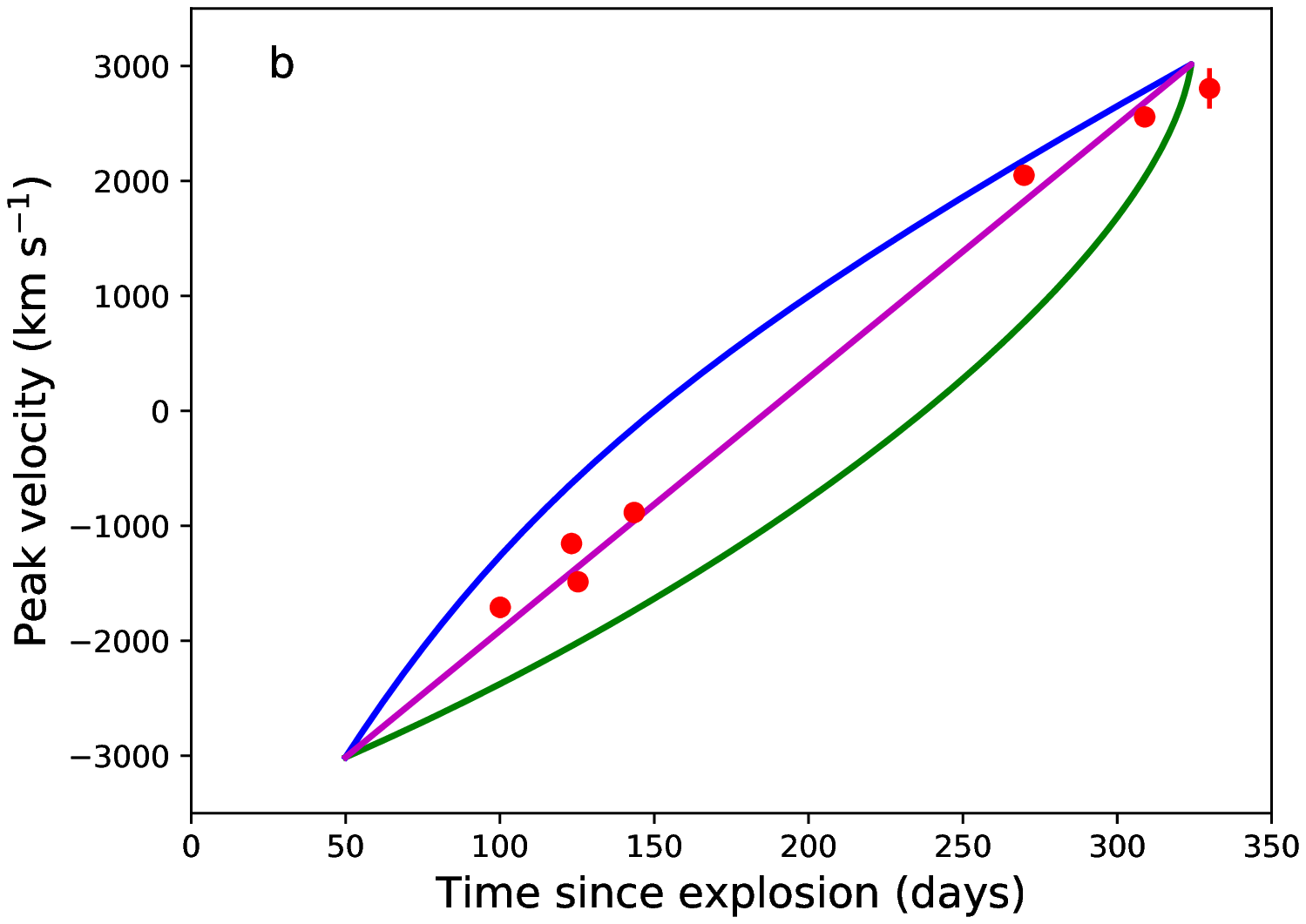}
\includegraphics[width=3.in]{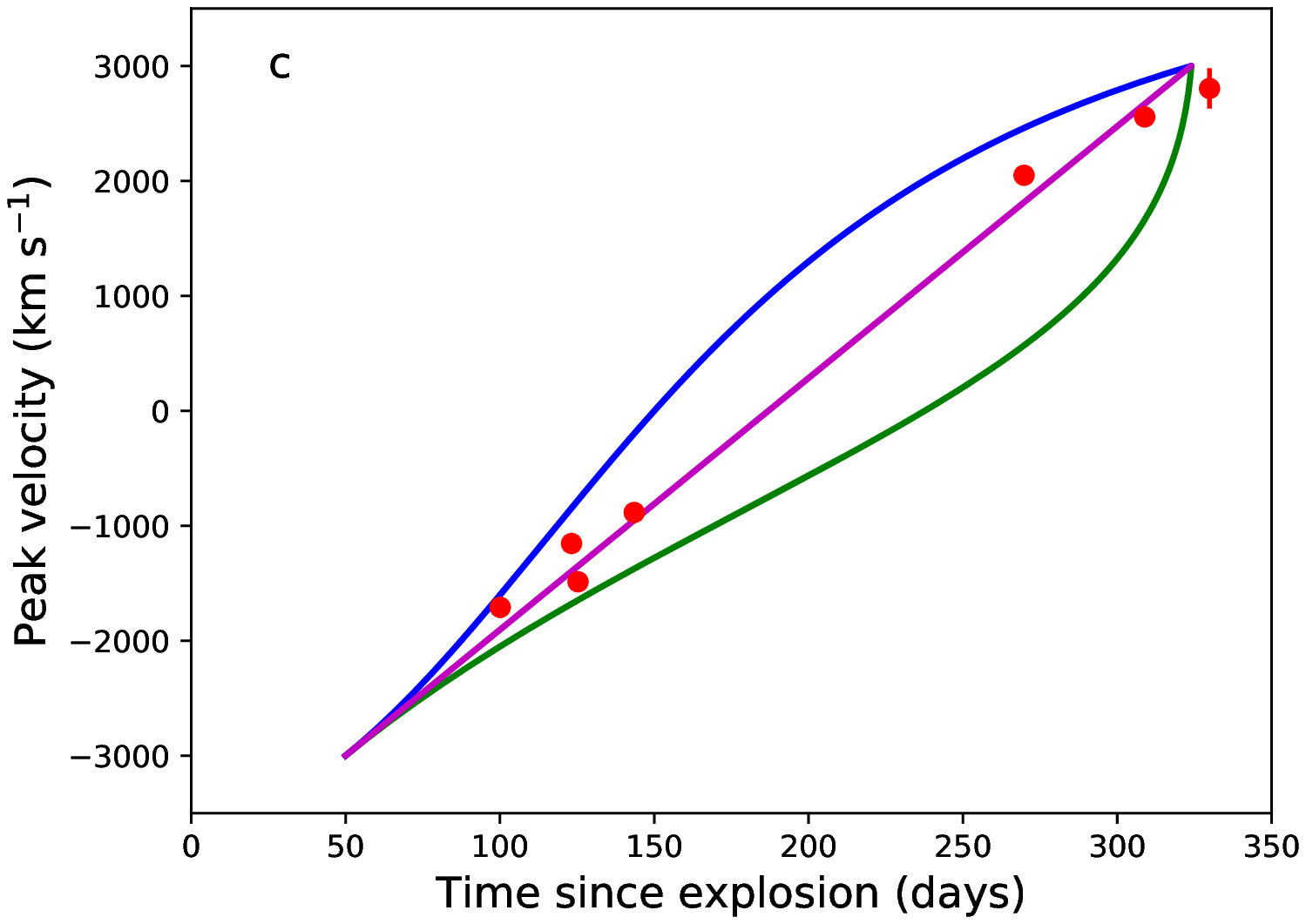}
\caption{Panel a: Same as Fig. \ref{fig:echo_parabola} but for two additional shell geometries, with ellipsoids with major (minor) axis 137 light days along the line of sight and minor (major) axis perpendicular to the line of sight and an axial ratio of 1.37. See text for details. 
Panels b and c: Peak velocity evolution for these geometries for a shell with $V\propto r$ (Panel b) and $V=$constant (Panel c). The purple line shows the spherical case, the blue line the ellipsoid with major axis 137 light days, and the green line the ellipsoid with minor axis 137 light days. Note the departures from the linear evolution in all of the non-spherical cases. 
\label{fig:velocity_geometry}}
\end{SIfigure}


\clearpage

\begin{deluxetable}{lcccc}
\tablewidth{0pt}
\tabletypesize{\footnotesize}
\tablecaption{iPTF16eh Photometry}
\tablehead{
\colhead{MJD} & 
\colhead{Rest-frame Phase} & 
\colhead{Filter} &
\colhead{AB Mag} &
\colhead{Instrument} \\
\colhead{(days)} &
\colhead{(days)}  & 
\colhead{} &
\colhead{} &
\colhead{} 
}
\startdata
57358.53 & $   -8.39$ &    g & $>20.71$ & P48 \\
57358.56 & $   -8.37$ &    g & $>20.86$ & P48 \\
57361.54 & $   -6.28$ &    g & $>20.64$ & P48 \\
57373.54 & $    2.13$ &    g & $21.43 \pm 0.23$ & P48 \\
57373.57 & $    2.15$ &    g & $>20.03$ & P48 \\
57388.55 & $   12.65$ &    g & $19.62 \pm 0.07$ & P48 \\
57388.58 & $   12.67$ &    g & $19.70 \pm 0.12$ & P48 \\
57400.55 & $   21.06$ &    g & $19.24 \pm 0.07$ & P48 \\
57420.53 & $   35.05$ &    g & $19.00 \pm 0.03$ & P48 \\
57420.56 & $   35.08$ &    g & $18.99 \pm 0.06$ & P48 
\enddata
\label{tab:phot}
\tablecomments{Full table is available as a separate, machine-readable file. A portion is shown here for clarity.}
\end{deluxetable}

\begin{deluxetable}{lccccc}
\tablewidth{0pt}
\tabletypesize{\footnotesize}
\tablecaption{Summary of Spectroscopic Observations}
\tablehead{
\colhead{Observation Date} & 
\colhead{Phase\tablenotemark{a}} &
\colhead{Telescope+Instrument} &
\colhead{Grating\tablenotemark{b}} &
\colhead{Exp. time\tablenotemark{b}} &
\colhead{Airmass} \\
\colhead{(YYYY MM DD.D)}  & 
\colhead{(rest-frame days)} &
\colhead{} &
\colhead{} &
\colhead{(s)} &
\colhead{} 
}
\startdata
2016 Feb 18.5 & $45.9$ & Subaru+FOCAS & \nodata & 600 & 1.03 \\
2016 Feb 27.5 & $52.5$ & P200+DBSP    & 600/4000, 316/7500  & 900 & 1.16 \\
2016 Mar 06.6 & $58.2$ & Keck I+LRIS   & 400/3400, 400/8500  & 280, 240 & 1.27 \\
2016 Apr 10.3 & $82.5$ & Keck I+LRIS   & 400/3400, 400/8500 & 300, 300 & 1.11 \\
2016 May 05.4 & $100.1$ & Keck I+LRIS   & 400/3400, 400/8500  & 1200, 1120 & 1.16 \\
2016 Jun 07.3 & $123.2$ & Keck I+LRIS   & 400/3400, 400/8500  & 600, 520 & 1.18 \\
2016 Jun 10.4 & $125.3$ & Keck I+LRIS   & 600/4000, 400/8500  & 1905, 1800 & 1.55 \\
2016 Jul 06.3 & $143.5$ & Keck I+LRIS   & 400/3400, 400/8500 & 2400, 2280 & 1.64 \\
2017 Jan 02.6 & $269.8$ & Keck I+LRIS   & 400/3400, 400/8500  & 2730, 2550 & 1.26 \\
2017 Feb 27.5 & $309.0$ & Keck I+LRIS   & 400/3400, 400/8500  & 3600, 3510 & 1.06 \\
2017 Mar 29.4 & $329.9$ & Keck I+LRIS   & 400/3400, 400/8500  & 5415, 5070 & 1.09 \\
2017 Apr 29.4 & $351.7$ & Keck I+LRIS   & 400/3400, 400/8500  & 7200, 7020 & 1.25
\enddata
\tablenotetext{a}{Relative to best-fit explosion date on 2015 December 14.5. To get phase relative to $g$-band peak, subtract 37.7~days.}
\tablenotetext{b}{Comma-separated values indicate setup for blue and red arms, respectively.}
\label{tab:spec}
\end{deluxetable}

\begin{deluxetable}{lcccc}
\tablewidth{0pt}
\tabletypesize{\footnotesize}
\tablecaption{Emission Line Properties}
\tablehead{
\colhead{MJD} & 
\colhead{Time since explosion} &
\colhead{Line centroid} &
\colhead{Line FWHM} &
\colhead{Line flux}  \\
\colhead{(days)}  & 
\colhead{(rest-frame days)} &
\colhead{(\AA\,, rest-frame)} &
\colhead{(\AA\,, rest-frame)} &
\colhead{($10^{-16}$~erg~s$^{-1}$~cm$^{-2}$)} 
}
\startdata
57513.4 & 100.1 & 2785.41 $\pm$ 0.85 & 13.7 $\pm$ 2.0 & 1.89 $\pm$ 0.37 \\
57546.3 & 123.2 & 2790.59 $\pm$ 0.76 & 15.7 $\pm$ 1.8 & 1.42 $\pm$ 0.22 \\
57549.4 & 125.3 & 2787.48 $\pm$ 0.73 & 12.0 $\pm$ 1.7 & 1.17 $\pm$ 0.23 \\
57575.3 & 143.5 & 2793.11 $\pm$ 0.67 & 16.6 $\pm$ 1.5 & 1.24 $\pm$ 0.16 \\
57755.6 & 269.8 & 2820.50 $\pm$ 0.24 & 13.7 $\pm$ 0.6 & 2.02 $\pm$ 0.11 \\
57811.5 & 309.0 & 2825.24 $\pm$ 0.29 & 12.5 $\pm$ 0.7 & 1.03 $\pm$ 0.07 \\
57841.4 & 329.9 & 2827.56 $\pm$ 1.64 & 14.4 $\pm$ 4.0 & 0.49 $\pm$ 0.17 \\
57872.4 & 351.7 &\nodata & \nodata & $<0.28$ 
\enddata
\tablecomments{Table is also available in machine-readable form.}
\end{deluxetable}

\begin{deluxetable}{lcccc}
\tablewidth{0pt}
\tabletypesize{\footnotesize}
\tablecaption{Blackbody Parameters and Pseudo-Bolometric Light Curve}
\tablehead{
\colhead{MJD} &
\colhead{Time since explosion} &
\colhead{Pseudo-bolometric luminosity} &
\colhead{Blackbody temperature} &
\colhead{Blackbody radius} \\
\colhead{(days)} &
\colhead{(rest-frame days)} &
\colhead{($10^{44}~{\rm erg~s}^{-1}$)} &
\colhead{($10^{3}$~K} &
\colhead{($10^{15}$~cm)}
}
\startdata
57427.3 & 39.8 & 2.73 $\pm$ 0.14 & 16.0 $\pm$ 1.4 & 3.6 $\pm$ 0.4 \\
57429.2 & 41.2 & 2.61 $\pm$ 0.07 & 15.6 $\pm$ 0.7 & 3.7 $\pm$ 0.2 \\
57431.2 & 42.6 & 2.63 $\pm$ 0.05 & 15.1 $\pm$ 0.5 & 3.9 $\pm$ 0.2 \\
57435.3 & 45.4 & 2.55 $\pm$ 0.07 & 14.2 $\pm$ 0.5 & 4.1 $\pm$ 0.2 \\
57439.2 & 48.1 & 2.67 $\pm$ 0.13 & 16.3 $\pm$ 1.5 & 3.5 $\pm$ 0.4 \\
57441.2 & 49.5 & 2.81 $\pm$ 0.23 & 12.6 $\pm$ 1.1 & 5.0 $\pm$ 0.7 \\
57443.2 & 50.9 & 2.53 $\pm$ 0.13 & 13.9 $\pm$ 1.0 & 4.2 $\pm$ 0.4 \\
57445.4 & 52.5 & 2.50 $\pm$ 0.09 & 13.0 $\pm$ 0.6 & 4.7 $\pm$ 0.3 \\
57451.5 & 56.7 & 2.66 $\pm$ 0.20 & 13.0 $\pm$ 1.3 & 4.8 $\pm$ 0.7 \\
57464.2 & 65.6 & 2.33 $\pm$ 0.08 & 12.3 $\pm$ 0.5 & 4.8 $\pm$ 0.3 
\enddata
\tablecomments{Full table is available as a separate, machine-readable file. A portion is shown here for clarity.}
\end{deluxetable}

\end{addendum}

\end{document}